\documentclass[11pt]{article}
\usepackage[dvips]{graphicx}
\usepackage{epsfig}
\usepackage[left]{lineno}
\usepackage{amsmath}
\usepackage{amssymb}
\usepackage{color}
\usepackage{hyperref}
\usepackage{adjustbox}
\usepackage{xspace}
\usepackage{multirow}

\usepackage{booktabs}
\usepackage{xltabular}
\usepackage{relsize}

\usepackage{alphalph}

\usepackage{float}

\definecolor{darkgreen}{rgb}{0.23, 0.5, 0.23}
\definecolor{amber}{rgb}{0.83, 0.67, 0}

\def\kpillc{K^\pm \to \pi^\pm \ell^+ \ell^-}

\def\kpills{K_{\pi \ell \ell}}

\def\kpimumu{K^+ \to \pi^+ \mu^+ \mu^-}
\def\kpimumus{K_{\pi \mu \mu}}

\def\kpiees{K_{\pi e e}}
\def\ke2{K^+ \to e^+ \nu_e}

\def\kmu2{K^+ \to \mu^+ \nu_\mu}

\def\ktpil{K^+ \to \pi^+ \pi^+ \pi^-}
\def\ktpis{K_{3\pi}}

\def\pimumug{\pi\mu\mu}

\def\geant {\mbox{\textsc{Geant4}}\xspace}

\let\oldequation\equation
\let\oldendequation\endequation
\renewenvironment{equation}
    {\linenomathNonumbers\oldequation}
    {\oldendequation\endlinenomath}

\newcommand{\appropto}{\mathrel{\vcenter{
  \offinterlineskip\halign{\hfil$##$\cr
    \propto\cr\noalign{\kern2pt}\sim\cr\noalign{\kern-2pt}}}}}

\begin{document}
\centerline{\LARGE EUROPEAN ORGANIZATION FOR NUCLEAR RESEARCH}

\vspace{10mm}
{\flushright{
 CERN-EP-2022-189 \\
 10 September 2022\\
}}
\vspace{10mm}

\begin{center}
\boldmath
{\bf {\Large\boldmath{A measurement of the $\kpimumu$ decay}}}
\unboldmath
\end{center}

\vspace{5mm}
\begin{center}
{\Large The NA62 Collaboration}
\end{center}

\begin{abstract}
A sample of $2.8 \times 10^4$ $\kpimumu$ candidates with negligible background was collected by the NA62 experiment at the CERN SPS in 2017--2018. The model-independent branching fraction is measured to be $(9.15 \pm 0.08)\times 10^{-8}$, a factor three more precise than previous measurements. The decay form factor is presented as a function of the squared dimuon mass. A measurement of the form factor parameters and their uncertainties is performed using a description based on Chiral Perturbation Theory at $\mathcal{O}(p^6)$.

\end{abstract}

\vspace{20mm}

\begin{center}
    \textit{Accepted for publication in JHEP}
\end{center}

\newpage
\begin{center}
{\Large The NA62 Collaboration\renewcommand{\thefootnote}{\fnsymbol{footnote}} %
\footnotemark[1]\renewcommand{\thefootnote}{\arabic{footnote}}}\\
\end{center}
\vspace{5mm}
{\flushleft{
 E.~Cortina Gil\footnotemark[1],
 A.~Kleimenova\footnotemark[1]$^,$\renewcommand{\thefootnote}{\alphalph{\value{footnote}}}\footnotemark[1]\renewcommand{\thefootnote}{\arabic{footnote}},
E.~Minucci\footnotemark[1]$^,$\renewcommand{\thefootnote}{\alphalph{\value{footnote}}}\footnotemark[2]$^,$\footnotemark[3]\renewcommand{\thefootnote}{\arabic{footnote}},
 S.~Padolski\footnotemark[1]$^,$\renewcommand{\thefootnote}{\alphalph{\value{footnote}}}\footnotemark[4]\renewcommand{\thefootnote}{\arabic{footnote}},
 P.~Petrov\footnotemark[1],
 A.~Shaikhiev\footnotemark[1]$^,$\renewcommand{\thefootnote}{\alphalph{\value{footnote}}}\footnotemark[5]\renewcommand{\thefootnote}{\arabic{footnote}},
 R.~Volpe\footnotemark[1]$^,$\renewcommand{\thefootnote}{\alphalph{\value{footnote}}}\footnotemark[6]\renewcommand{\thefootnote}{\arabic{footnote}},  
 T.~Numao\footnotemark[2],
 Y.~Petrov\footnotemark[2],
 B.~Velghe\footnotemark[2],
 V.W.S.~Wong\footnotemark[2], 
 D.~Bryman\footnotemark[3]$^,$\renewcommand{\thefootnote}{\alphalph{\value{footnote}}}\footnotemark[7]\renewcommand{\thefootnote}{\arabic{footnote}},
 J.~Fu\footnotemark[3], 
 T.~Husek\footnotemark[4]$^,$\renewcommand{\thefootnote}{\alphalph{\value{footnote}}}\footnotemark[8]\renewcommand{\thefootnote}{\arabic{footnote}},
 J.~Jerhot\footnotemark[4]$^,$\renewcommand{\thefootnote}{\alphalph{\value{footnote}}}\footnotemark[9]\renewcommand{\thefootnote}{\arabic{footnote}},
 K.~Kampf,\footnotemark[4],
 M.~Zamkovsky\footnotemark[4]$^,$\renewcommand{\thefootnote}{\alphalph{\value{footnote}}}\footnotemark[2]\renewcommand{\thefootnote}{\arabic{footnote}},
 R.~Aliberti\footnotemark[5]$^,$\renewcommand{\thefootnote}{\alphalph{\value{footnote}}}\footnotemark[10]\renewcommand{\thefootnote}{\arabic{footnote}},
 G.~Khoriauli\footnotemark[5]$^,$\renewcommand{\thefootnote}{\alphalph{\value{footnote}}}\footnotemark[11]\renewcommand{\thefootnote}{\arabic{footnote}},
 J.~Kunze\footnotemark[5],
 D.~Lomidze\footnotemark[5]$^,$\renewcommand{\thefootnote}{\alphalph{\value{footnote}}}\footnotemark[12]\renewcommand{\thefootnote}{\arabic{footnote}},
 L.~Peruzzo\footnotemark[5],
 M.~Vormstein\footnotemark[5],
 R.~Wanke\footnotemark[5], 
 P.~Dalpiaz\footnotemark[6],
 M.~Fiorini\footnotemark[6],
 I.~Neri\footnotemark[6],
 \mbox{A.~Norton\footnotemark[6]$^,$\renewcommand{\thefootnote}{\alphalph{\value{footnote}}}\footnotemark[13]\renewcommand{\thefootnote}{\arabic{footnote}},}
 F.~Petrucci\footnotemark[6],
 H.~Wahl\footnotemark[6]$^,$\renewcommand{\thefootnote}{\alphalph{\value{footnote}}}\footnotemark[14]\renewcommand{\thefootnote}{\arabic{footnote}},  
 \mbox{A.~Cotta Ramusino\footnotemark[7],}
 A.~Gianoli\footnotemark[7],  
 E.~Iacopini\footnotemark[8],
 G.~Latino\footnotemark[8],
 M.~Lenti\footnotemark[8],
 A.~Parenti\footnotemark[8], 
 A.~Bizzeti\footnotemark[9]$^,$\renewcommand{\thefootnote}{\alphalph{\value{footnote}}}\footnotemark[15]\renewcommand{\thefootnote}{\arabic{footnote}}, 
 F.~Bucci\footnotemark[9], 
 A.~Antonelli\footnotemark[10],
G.~Georgiev\footnotemark[10]$^,$\renewcommand{\thefootnote}{\alphalph{\value{footnote}}}\footnotemark[16]\renewcommand{\thefootnote}{\arabic{footnote}}, 
\mbox{V.~Kozhuharov\footnotemark[10]$^,$\renewcommand{\thefootnote}{\alphalph{\value{footnote}}}\footnotemark[16]\renewcommand{\thefootnote}{\arabic{footnote}},} 
G.~Lanfranchi\footnotemark[10],
S.~Martellotti\footnotemark[10],
M.~Moulson\footnotemark[10],
T.~Spadaro\footnotemark[10],
G.~Tinti\footnotemark[10], 
 F.~Ambrosino\footnotemark[11],
 T.~Capussela\footnotemark[11],
 M.~Corvino\footnotemark[11]$^,$\renewcommand{\thefootnote}{\alphalph{\value{footnote}}}\footnotemark[2]\renewcommand{\thefootnote}{\arabic{footnote}}, 
 D.~Di Filippo\footnotemark[11],
 R.~Fiorenza\footnotemark[11]$^,$\renewcommand{\thefootnote}{\alphalph{\value{footnote}}}\footnotemark[17]\renewcommand{\thefootnote}{\arabic{footnote}}, 
 P.~Massarotti\footnotemark[11],
 M.~Mirra\footnotemark[11],
 M.~Napolitano\footnotemark[11],
 G.~Saracino\footnotemark[11],  
 G.~Anzivino\footnotemark[12],
 F.~Brizioli\footnotemark[12]$^,$\renewcommand{\thefootnote}{\alphalph{\value{footnote}}}\footnotemark[2]\renewcommand{\thefootnote}{\arabic{footnote}}, 
 E.~Imbergamo\footnotemark[12],
 R.~Lollini\footnotemark[12],
 R.~Piandani\footnotemark[12]$^,$\renewcommand{\thefootnote}{\alphalph{\value{footnote}}}\footnotemark[18]\renewcommand{\thefootnote}{\arabic{footnote}}, 
 C.~Santoni\footnotemark[12],  
 M.~Barbanera\footnotemark[13],
 P.~Cenci\footnotemark[13],
 B.~Checcucci\footnotemark[13],
 P.~Lubrano\footnotemark[13],
 M.~Lupi\footnotemark[13]$^,$\renewcommand{\thefootnote}{\alphalph{\value{footnote}}}\footnotemark[19]\renewcommand{\thefootnote}{\arabic{footnote}}, 
 M.~Pepe\footnotemark[13],
 M.~Piccini\footnotemark[13],  
{
 F.~Costantini\footnotemark[14],
L.~Di Lella\footnotemark[14]$^,$\renewcommand{\thefootnote}{\alphalph{\value{footnote}}}\footnotemark[14]\renewcommand{\thefootnote}{\arabic{footnote}}, 
 N.~Doble\footnotemark[14]$^,$\renewcommand{\thefootnote}{\alphalph{\value{footnote}}}\footnotemark[14]\renewcommand{\thefootnote}{\arabic{footnote}}, 
 M.~Giorgi\footnotemark[14],
 S.~Giudici\footnotemark[14],
 G.~Lamanna\footnotemark[14],
 E.~Lari\footnotemark[14],
 E.~Pedreschi\footnotemark[14],
 M.~Sozzi\footnotemark[14],  
 C.~Cerri\footnotemark[15],
 R.~Fantechi\footnotemark[15],
 L.~Pontisso\footnotemark[15]$^,$\renewcommand{\thefootnote}{\alphalph{\value{footnote}}}\footnotemark[20]\renewcommand{\thefootnote}{\arabic{footnote}}, 
 F.~Spinella\footnotemark[15],  
 I.~Mannelli\footnotemark[16],   
 G.~D'Agostini\footnotemark[17], 
 M.~Raggi\footnotemark[17],  
 A.~Biagioni\footnotemark[18], 
 P.~Cretaro\footnotemark[18], 
 O.~Frezza\footnotemark[18], 
 E.~Leonardi\footnotemark[18], 
 A.~Lonardo\footnotemark[18], 
 M.~Turisini\footnotemark[18], 
 P.~Valente\footnotemark[18], 
 P.~Vicini\footnotemark[18],  
 R.~Ammendola\footnotemark[19], 
 V.~Bonaiuto\footnotemark[19]$^,$\renewcommand{\thefootnote}{\alphalph{\value{footnote}}}\footnotemark[21]\renewcommand{\thefootnote}{\arabic{footnote}}, 
 A.~Fucci\footnotemark[19], 
 A.~Salamon\footnotemark[19], 
 F.~Sargeni\footnotemark[19]$^,$\renewcommand{\thefootnote}{\alphalph{\value{footnote}}}\footnotemark[22]\renewcommand{\thefootnote}{\arabic{footnote}},   
 R.~Arcidiacono\footnotemark[20]$^,$\renewcommand{\thefootnote}{\alphalph{\value{footnote}}}\footnotemark[23]\renewcommand{\thefootnote}{\arabic{footnote}}, 
 B.~Bloch-Devaux\footnotemark[20],
 M.~Boretto\footnotemark[20]$^,$\renewcommand{\thefootnote}{\alphalph{\value{footnote}}}\footnotemark[2]\renewcommand{\thefootnote}{\arabic{footnote}}, 
 E.~Menichetti\footnotemark[20],
 E.~Migliore\footnotemark[20],
 D.~Soldi\footnotemark[20],   
 C.~Biino\footnotemark[21],
 A.~Filippi\footnotemark[21],
 F.~Marchetto\footnotemark[21],  
 J.~Engelfried\footnotemark[22],
 N.~Estrada-Tristan\footnotemark[22]$^,$\renewcommand{\thefootnote}{\alphalph{\value{footnote}}}\footnotemark[24]\renewcommand{\thefootnote}{\arabic{footnote}},  
 A.M.~Bragadireanu\footnotemark[23],
 S.A.~Ghinescu\footnotemark[23],
 O.E.~Hutanu\footnotemark[23], 
 A.~Baeva\footnotemark[24],
 D.~Baigarashev\footnotemark[24]$^,$\renewcommand{\thefootnote}{\alphalph{\value{footnote}}}\footnotemark[25]\renewcommand{\thefootnote}{\arabic{footnote}}, 
 D.~Emelyanov\footnotemark[24],
 T.~Enik\footnotemark[24],
 V.~Falaleev\footnotemark[24]$^,$\renewcommand{\thefootnote}{\alphalph{\value{footnote}}}\footnotemark[26]\renewcommand{\thefootnote}{\arabic{footnote}}, 
 V.~Kekelidze\footnotemark[24],
 A.~Korotkova\footnotemark[24],
 L.~Litov\footnotemark[24]$^,$\renewcommand{\thefootnote}{\alphalph{\value{footnote}}}\footnotemark[16]\renewcommand{\thefootnote}{\arabic{footnote}}, 
 D.~Madigozhin\footnotemark[24],
 M.~Misheva\footnotemark[24]$^,$\renewcommand{\thefootnote}{\alphalph{\value{footnote}}}\footnotemark[27]\renewcommand{\thefootnote}{\arabic{footnote}}, 
 N.~Molokanova\footnotemark[24],
 S.~Movchan\footnotemark[24],
 I.~Polenkevich\footnotemark[24],
 Yu.~Potrebenikov\footnotemark[24],
 S.~Shkarovskiy\footnotemark[24],
 A.~Zinchenko\footnotemark[24]$^,$\renewcommand{\thefootnote}{\fnsymbol{footnote}}\footnotemark[2]\renewcommand{\thefootnote}{\arabic{footnote}},  
 S.~Fedotov\footnotemark[25],
 E.~Gushchin\footnotemark[25],
 A.~Khotyantsev\footnotemark[25],
 Y.~Kudenko\footnotemark[25]$^,$\renewcommand{\thefootnote}{\alphalph{\value{footnote}}}\footnotemark[28]\renewcommand{\thefootnote}{\arabic{footnote}},  
 V.~Kurochka\footnotemark[25],
 M.~Medvedeva\footnotemark[25],
 A.~Mefodev \footnotemark[25], 
 S.~Kholodenko\footnotemark[26],
 V.~Kurshetsov\footnotemark[26],
 V.~Obraztsov\footnotemark[26],
 A.~Ostankov\footnotemark[26]$^,$\renewcommand{\thefootnote}{\fnsymbol{footnote}}\footnotemark[2]\renewcommand{\thefootnote}{\arabic{footnote}},
 V.~Semenov\footnotemark[26]$^,$\renewcommand{\thefootnote}{\fnsymbol{footnote}}\footnotemark[2]\renewcommand{\thefootnote}{\arabic{footnote}},
 V.~Sugonyaev\footnotemark[26],
 O.~Yushchenko\footnotemark[26],  
 L.~Bician\footnotemark[27]$^,$\renewcommand{\thefootnote}{\alphalph{\value{footnote}}}\footnotemark[29]\renewcommand{\thefootnote}{\arabic{footnote}}$^,$\renewcommand{\thefootnote}{\fnsymbol{footnote}}\footnotemark[1]\renewcommand{\thefootnote}{\arabic{footnote}}, 
 T.~Blazek\footnotemark[27],
 V.~Cerny\footnotemark[27],
 Z.~Kucerova\footnotemark[27],  
 J.~Bernhard\footnotemark[28],
 A.~Ceccucci\footnotemark[28],
 H.~Danielsson\footnotemark[28],
 N.~De Simone\footnotemark[28]$^,$\renewcommand{\thefootnote}{\alphalph{\value{footnote}}}\footnotemark[30]\renewcommand{\thefootnote}{\arabic{footnote}},
 F.~Duval\footnotemark[28],
 B.~D\"obrich\footnotemark[28],
 L.~Federici\footnotemark[28],
 E.~Gamberini\footnotemark[28],
 L.~Gatignon\footnotemark[28]$^,$\renewcommand{\thefootnote}{\alphalph{\value{footnote}}}\footnotemark[31]\renewcommand{\thefootnote}{\arabic{footnote}},
 R.~Guida\footnotemark[28],
 F.~Hahn\footnotemark[28]$^,$\renewcommand{\thefootnote}{\fnsymbol{footnote}}\footnotemark[2]\renewcommand{\thefootnote}{\arabic{footnote}},
 E.B.~Holzer\footnotemark[28],
 B.~Jenninger\footnotemark[28],
 M.~Koval\footnotemark[28]$^,$\renewcommand{\thefootnote}{\alphalph{\value{footnote}}}\footnotemark[29]\renewcommand{\thefootnote}{\arabic{footnote}},
 P.~Laycock\footnotemark[28]$^,$\renewcommand{\thefootnote}{\alphalph{\value{footnote}}}\footnotemark[4]\renewcommand{\thefootnote}{\arabic{footnote}},
 G.~Lehmann Miotto\footnotemark[28],
 P.~Lichard\footnotemark[28],
 A.~Mapelli\footnotemark[28],
 R.~Marchevski\footnotemark[28]$^,$\renewcommand{\thefootnote}{\alphalph{\value{footnote}}}\footnotemark[32]\renewcommand{\thefootnote}{\arabic{footnote}},
 K.~Massri\footnotemark[28],
 M.~Noy\footnotemark[28],
 V.~Palladino\footnotemark[28],
 M.~Perrin-Terrin\footnotemark[28]$^,$\renewcommand{\thefootnote}{\alphalph{\value{footnote}}}\footnotemark[33]$^,$\footnotemark[34]\renewcommand{\thefootnote}{\arabic{footnote}},
 J.~Pinzino\footnotemark[28]$^,$\renewcommand{\thefootnote}{\alphalph{\value{footnote}}}\footnotemark[35]\renewcommand{\thefootnote}{\arabic{footnote}}\renewcommand{\thefootnote}{\arabic{footnote}},
 V.~Ryjov\footnotemark[28],
 S.~Schuchmann\footnotemark[28],
 S.~Venditti\footnotemark[28],  
 T.~Bache\footnotemark[29],
 M.B.~Brunetti\footnotemark[29]$^,$\renewcommand{\thefootnote}{\alphalph{\value{footnote}}}\footnotemark[36]\renewcommand{\thefootnote}{\arabic{footnote}},
 V.~Duk\footnotemark[29]$^,$\renewcommand{\thefootnote}{\alphalph{\value{footnote}}}\footnotemark[6]\renewcommand{\thefootnote}{\arabic{footnote}},
 V.~Fascianelli\footnotemark[29]$^,$\renewcommand{\thefootnote}{\alphalph{\value{footnote}}}\footnotemark[37]\renewcommand{\thefootnote}{\arabic{footnote}},
 J. R.~Fry\footnotemark[29],
 F.~Gonnella\footnotemark[29],
 E.~Goudzovski\footnotemark[29],
 J.~Henshaw\footnotemark[29],
 L.~Iacobuzio\footnotemark[29],
 C.~Lazzeroni\footnotemark[29],
 N.~Lurkin\footnotemark[29]$^,$\renewcommand{\thefootnote}{\alphalph{\value{footnote}}}\footnotemark[9]\renewcommand{\thefootnote}{\arabic{footnote}},
 F.~Newson\footnotemark[29],
 C.~Parkinson\footnotemark[29], 
 A.~Romano\footnotemark[29], 
 A.~Sergi\footnotemark[29]$^,$\renewcommand{\thefootnote}{\alphalph{\value{footnote}}}\footnotemark[38]\renewcommand{\thefootnote}{\arabic{footnote}},
 A.~Sturgess\footnotemark[29],
 J.~Swallow\footnotemark[29]$^,$\renewcommand{\thefootnote}{\alphalph{\value{footnote}}}\footnotemark[2]\renewcommand{\thefootnote}{\arabic{footnote}},
 A.~Tomczak\footnotemark[29],  
 H.~Heath\footnotemark[30],
 R.~Page\footnotemark[30],
 S.~Trilov\footnotemark[30], 
 B.~Angelucci\footnotemark[31],
 D.~Britton\footnotemark[31],
 C.~Graham\footnotemark[31],
 D.~Protopopescu\footnotemark[31],  
 J.~Carmignani\footnotemark[32]$^,$\renewcommand{\thefootnote}{\alphalph{\value{footnote}}}\footnotemark[39]\renewcommand{\thefootnote}{\arabic{footnote}}, 
 J.B.~Dainton\footnotemark[32],
 R.W. L.~Jones\footnotemark[32],
 G.~Ruggiero\footnotemark[32]$^,$\renewcommand{\thefootnote}{\alphalph{\value{footnote}}}\footnotemark[40]\renewcommand{\thefootnote}{\arabic{footnote}},   
 L.~Fulton\footnotemark[33],
 D.~Hutchcroft\footnotemark[33],
 E.~Maurice\footnotemark[33]$^,$\renewcommand{\thefootnote}{\alphalph{\value{footnote}}}\footnotemark[41]\renewcommand{\thefootnote}{\arabic{footnote}},  
 B.~Wrona\footnotemark[33],  
 A.~Conovaloff\footnotemark[34],
 P.~Cooper\footnotemark[34],
 D.~Coward\footnotemark[34]$^,$\renewcommand{\thefootnote}{\alphalph{\value{footnote}}}\footnotemark[42]\renewcommand{\thefootnote}{\arabic{footnote}},  
 P.~Rubin\footnotemark[34]  
 \vspace{45mm}
} }  
\newlength{\basefootnotesep}
\setlength{\basefootnotesep}{\footnotesep}
\setcounter{footnote}{0}
\renewcommand{\thefootnote}{\fnsymbol{footnote}}
\footnotetext[1]{Corresponding author:  L.~Bician, email: lubos.bician@cern.ch}
\footnotetext[2]{Deceased}
\renewcommand{\thefootnote}{\arabic{footnote}}
\footnotetext[1]{Universit\'e Catholique de Louvain, B-1348 Louvain-La-Neuve, Belgium} 
%
\footnotetext[2]{TRIUMF, Vancouver, British Columbia, V6T 2A3, Canada} 
%
\footnotetext[3]{University of British Columbia, Vancouver, British Columbia, V6T 1Z4, Canada} 
%
\footnotetext[4]{Charles University, 116 36 Prague 1, Czech Republic} 
%
\footnotetext[5]{Institut f\"ur Physik and PRISMA Cluster of Excellence, Universit\"at Mainz, D-55099 Mainz, Germany} 
%
\footnotetext[6]{Dipartimento di Fisica e Scienze della Terra dell'Universit\`a e INFN, Sezione di Ferrara, I-44122 Ferrara, Italy} 
%
\footnotetext[7]{INFN, Sezione di Ferrara, I-44122 Ferrara, Italy} 
%
\footnotetext[8]{Dipartimento di Fisica e Astronomia dell'Universit\`a e INFN, Sezione di Firenze, I-50019 Sesto Fiorentino, Italy} 
%
\footnotetext[9]{INFN, Sezione di Firenze, I-50019 Sesto Fiorentino, Italy} 
%
\footnotetext[10]{Laboratori Nazionali di Frascati, I-00044 Frascati, Italy} 
%
\footnotetext[11]{Dipartimento di Fisica ``Ettore Pancini'' e INFN, Sezione di Napoli, I-80126 Napoli, Italy} 
%
\footnotetext[12]{Dipartimento di Fisica e Geologia dell'Universit\`a e INFN, Sezione di Perugia, I-06100 Perugia, Italy} 
%
\footnotetext[13]{INFN, Sezione di Perugia, I-06100 Perugia, Italy} 
%
\footnotetext[14]{Dipartimento di Fisica dell'Universit\`a e INFN, Sezione di Pisa, I-56100 Pisa, Italy} 
%
\footnotetext[15]{INFN, Sezione di Pisa, I-56100 Pisa, Italy} 
%
\footnotetext[16]{Scuola Normale Superiore e INFN, Sezione di Pisa, I-56100 Pisa, Italy} 
%
\footnotetext[17]{Dipartimento di Fisica, Sapienza Universit\`a di Roma e INFN, Sezione di Roma I, I-00185 Roma, Italy} 
%
\footnotetext[18]{INFN, Sezione di Roma I, I-00185 Roma, Italy} 
%
\footnotetext[19]{INFN, Sezione di Roma Tor Vergata, I-00133 Roma, Italy} 
%
\footnotetext[20]{Dipartimento di Fisica dell'Universit\`a e INFN, Sezione di Torino, I-10125 Torino, Italy} 
%
\footnotetext[21]{INFN, Sezione di Torino, I-10125 Torino, Italy} 
%
\footnotetext[22]{Instituto de F\'isica, Universidad Aut\'onoma de San Luis Potos\'i, 78240 San Luis Potos\'i, Mexico} 
%
\footnotetext[23]{Horia Hulubei National Institute for R\&D in Physics and Nuclear Engineering, 077125 Bucharest-Magurele, Romania} 
%
\footnotetext[24]
{Joint Institute for Nuclear Research, 141980 Dubna (MO), Russia} 
%
\footnotetext[25]
{Institute for Nuclear Research of the Russian Academy of Sciences, 117312 Moscow, Russia} 
%
\footnotetext[26]
{Institute for High Energy Physics of the Russian Federation, State Research Center ``Kurchatov Institute", 142281 Protvino (MO), Russia} 
%
\footnotetext[27]{Faculty of Mathematics, Physics and Informatics, Comenius University, 842 48, Bratislava, Slovakia} 
%
\footnotetext[28]{CERN,  European Organization for Nuclear Research, CH-1211 Geneva 23, Switzerland} 
%
\footnotetext[29]{University of Birmingham, Edgbaston, Birmingham, B15 2TT, UK} 
%
\footnotetext[30]{University of Bristol, Bristol, BS8 1TH, UK} 
%
\footnotetext[31]{University of Glasgow, Glasgow, G12 8QQ, UK} 
%
\footnotetext[32]{University of Lancaster, Lancaster, LA1 4YW, UK} 
%
\footnotetext[33]{University of Liverpool, Liverpool, L69 7ZE, UK} 
%
\footnotetext[34]{George Mason University, Fairfax, VA 22030, USA} 
\vspace{-40mm}
\renewcommand{\thefootnote}{\alphalph{\value{footnote}}}
\footnotetext[1]{Present address: Faculty of Mathematics, Physics and Informatics, Comenius University, 842 48, Bratislava, Slovakia}
\footnotetext[2]{Present address: CERN,  European Organization for Nuclear Research, CH-1211 Geneva 23, Switzerland}
\footnotetext[3]{Also at Laboratori Nazionali di Frascati, I-00044 Frascati, Italy}
\footnotetext[4]{Present address: Brookhaven National Laboratory, Upton, NY 11973, USA}
\footnotetext[5]{Present address: School of Physics and Astronomy, University of Birmingham, Birmingham, B15 2TT, UK}
\footnotetext[6]{Present address: INFN, Sezione di Perugia, I-06100 Perugia, Italy}
\footnotetext[7]{Also at TRIUMF, Vancouver, British Columbia, V6T 2A3, Canada}
\footnotetext[8]{Present address: Department of Astronomy and Theoretical Physics, Lund University, Lund, SE 223-62, Sweden}
\footnotetext[9]{Present address: Universit\'e Catholique de Louvain, B-1348 Louvain-La-Neuve, Belgium}
\footnotetext[10]{Present address: Institut f\"ur Kernphysik and Helmholtz Institute Mainz, Universit\"at Mainz, Mainz, D-55099, Germany}
\footnotetext[11]{Present address: Universit\"at W\"urzburg, D-97070 W\"urzburg, Germany}
\footnotetext[12]{Present address: European XFEL GmbH, D-22761 Hamburg, Germany}
\footnotetext[13]{Present address: University of Glasgow, Glasgow, G12 8QQ, UK}
\footnotetext[14]{Present address: Institut f\"ur Physik and PRISMA Cluster of Excellence, Universit\"at Mainz, D-55099 Mainz, Germany}
\footnotetext[15]{Also at Dipartimento di Scienze Fisiche, Informatiche e Matematiche, Universit\`a di Modena e Reggio Emilia, I-41125 Modena, Italy}
\footnotetext[16]{Also at Faculty of Physics, University of Sofia, BG-1164 Sofia, Bulgaria}
\footnotetext[17]{Present address: Scuola Superiore Meridionale e INFN, Sezione di Napoli, I-80138 Napoli, Italy}
\footnotetext[18]{Present address: Instituto de F\'isica, Universidad Aut\'onoma de San Luis Potos\'i, 78240 San Luis Potos\'i, Mexico}
\footnotetext[19]{Present address: Institut am Fachbereich Informatik und Mathematik, Goethe Universit\"at, D-60323 Frankfurt am Main, Germany}
\footnotetext[20]{Present address: INFN, Sezione di Roma I, I-00185 Roma, Italy}
\footnotetext[21]{Also at Department of Industrial Engineering, University of Roma Tor Vergata, I-00173 Roma, Italy}
\footnotetext[22]{Also at Department of Electronic Engineering, University of Roma Tor Vergata, I-00173 Roma, Italy}
\footnotetext[23]{Also at Universit\`a degli Studi del Piemonte Orientale, I-13100 Vercelli, Italy}
\footnotetext[24]{Also at Universidad de Guanajuato, 36000 Guanajuato, Mexico}
\footnotetext[25]{Also at L.N. Gumilyov Eurasian National University, 010000 Nur-Sultan, Kazakhstan}
\footnotetext[26]
{Also at Institute for Nuclear Research of the Russian Academy of Sciences, 117312 Moscow, Russia}
\footnotetext[27]{Present address: Institute of Nuclear Research and Nuclear Energy of Bulgarian Academy of Science (INRNE-BAS), BG-1784 Sofia, Bulgaria}
\footnotetext[28]
{Also at National Research Nuclear University (MEPhI), 115409 Moscow and Moscow Institute of Physics and Technology, 141701 Moscow region, Moscow, Russia}
\footnotetext[29]{Present address: Charles University, 116 36 Prague 1, Czech Republic}
\footnotetext[30]{Present address: DESY, D-15738 Zeuthen, Germany}
\footnotetext[31]{Present address: University of Lancaster, Lancaster, LA1 4YW, UK}
\footnotetext[32]{Present address: Weizmann Institute, Rehovot, 76100, Israel}
\footnotetext[33]{Present address: Aix Marseille University, CNRS/IN2P3, CPPM, F-13288, Marseille, France}
\footnotetext[34]{Also at Universit\'e Catholique de Louvain, B-1348 Louvain-La-Neuve, Belgium}
\footnotetext[35]{Present address: INFN, Sezione di Pisa, I-56100 Pisa, Italy}
\footnotetext[36]{Present address: Department of Physics, University of Warwick, Coventry, CV4 7AL, UK}
\footnotetext[37]{Present address: Center for theoretical neuroscience, Columbia University, New York, NY 10027, USA}
\footnotetext[38]{Present address: Dipartimento di Fisica dell'Universit\`a e INFN, Sezione di Genova, I-16146 Genova, Italy}
\footnotetext[39]{Present address: University of Liverpool, Liverpool, L69 7ZE, UK}
\footnotetext[40]{Present address: Dipartimento di Fisica e Astronomia dell'Universit\`a e INFN, Sezione di Firenze, I-50019 Sesto Fiorentino, Italy}
\footnotetext[41]{Present address: Laboratoire Leprince Ringuet, F-91120 Palaiseau, France}
\footnotetext[42]{Also at SLAC National Accelerator Laboratory, Stanford University, Menlo Park, CA 94025, USA}

}
\clearpage

\section*{Introduction}
\label{sec:intro}

The flavour-changing neutral current decays $\kpillc$ (denoted $\kpills$), with $\ell = e, \mu$ have been the focus of extensive theoretical work~\cite{Ecker:1987, D'Ambrosio:1998, Dubnickova:2006, Kubis:2010}. Dominant contributions to the $\kpills$ decays are mediated by virtual photon exchange $K^\pm \to \pi^\pm \gamma^* \to \pi^\pm \ell^+ \ell^-$ and involve long-distance hadronic effects described by a vector interaction form factor.

Studies of the $\kpiees$ and $\kpimumus$ decay form factors contribute to experimental tests of lepton flavour universality~\cite{Tunstall:2016bal, LFUV-kaons}. The first lattice QCD calculation of the form factor value at a specific lepton pair mass (lying outside the $\kpimumus$ kinematic region) using physical light-quark masses is presented in~\cite{LatticeQCD-Kpmm}. Future methodology optimizations together with advances in computing technology are expected to provide competitive lattice QCD predictions of the form factor.

The E787 collaboration at the Brookhaven National Laboratory reported the first observation of the $\kpimumus$ decay in 1997~\cite{E787-pmm}, which was followed by the E865~\cite{E865-pmm} and HyperCP~\cite{HyperCP} measurements. The E865 result established the vector nature of the decay form factor, while HyperCP studied both $\kpimumus^+$ and $\kpimumus^-$ decays and measured the CP violating decay rate asymmetry, found to be compatible with zero. The most precise study~\cite{NA48/2-pimumu} of $\kpimumus$ was performed by the NA48/2 collaboration at the CERN SPS. The $\kpiees$ decay was first observed at the CERN PS by the Geneva-Saclay collaboration in 1975~\cite{CERN-piee}, and subsequently measured by the E777~\cite{E777-piee}, E865~\cite{E865-piee} and NA48/2~\cite{NA48/2-piee} experiments. A summary of form factor measurements can be found in~\cite{FitOfBeta}.


Improved measurements of the $\kpimumus$ model-independent branching fraction and form factor parameters, based on the dataset collected in 2017--2018 by the NA62 experiment at the CERN SPS, are presented in the following. The forward-backward asymmetry of the decay with respect to angle $\theta_{K\mu}$ between the $K^+$ and the $\mu^-$ three-momenta in the $\mu^+\mu^-$ rest frame, is also measured.

\section{Beam, detector and data sample}
\label{sec:beam_det_sample}

\begin{figure}[t]
\begin{center}
\resizebox{\textwidth}{!}{\includegraphics{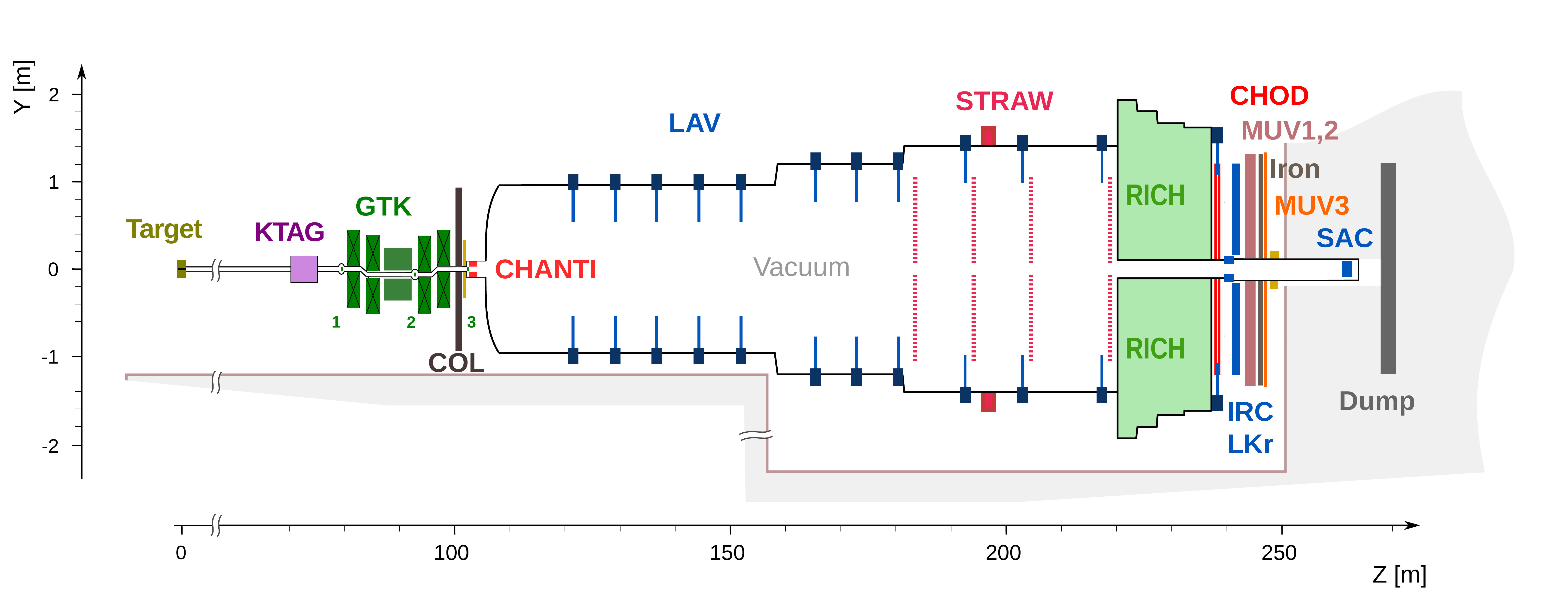}}
\put(-343,67){\scriptsize\color{darkgreen}\rotatebox{90}{\textbf{\textsf{SCR}}}}
\put(-320,105){\tiny\color{amber}\rotatebox{90}{\textbf{\textsf{TRIM5}}}}
\put(-172,46){\scriptsize\color{red}{\textbf{\textsf{M}}}}
\end{center}
\vspace{-12mm}
\caption{Schematic side view of the NA62 detector.}
\label{fig:detector}
\end{figure}

The layout of the NA62 beamline and detector~\cite{DetectorPaper} is shown schematically in \autoref{fig:detector}. An unseparated secondary beam of $\pi^+$ (70\%), protons (23\%) and $K^+$ (6\%) is created by directing 400~GeV/$c$ protons extracted from the CERN SPS onto a beryllium target in spills of 3~s effective duration. The target position defines the origin of the NA62 reference system: the beam travels along the $Z$ axis in the positive direction (downstream), the $Y$ axis points vertically up, and the $X$ axis is horizontal and directed to form a right-handed coordinate  system. The central beam momentum is 75~GeV/$c$, with a momentum spread of 1\% (rms).

Beam kaons are tagged with a time resolution of 70~ps by a differential Cherenkov counter (KTAG), which uses nitrogen gas at 1.75~bar pressure contained in a 5~m long vessel as radiator. Beam particle positions, momenta and times (to better than 100~ps resolution) are measured by a silicon pixel spectrometer consisting of three stations (GTK1,2,3) and four dipole magnets. A~toroidal muon sweeper, called scraper (SCR), is installed between GTK1 and GTK2. A 1.2~m thick steel collimator (COL) with a $76\times40$~mm$^2$ central aperture and $1.7\times1.8$~m$^2$ outer dimensions is placed upstream of GTK3 to absorb hadrons from upstream $K^+$ decays; a variable aperture collimator of \mbox{$0.15\times0.15$~m$^2$} outer dimensions was used up to early 2018. Inelastic interactions of beam particles in GTK3 are detected by an array of scintillator hodoscopes (CHANTI). A dipole magnet (TRIM5) providing a 90 MeV$/c$ horizontal momentum kick is located in front of GTK3. The beam is delivered into a vacuum tank evacuated to a pressure of $10^{-6}$~mbar, which contains a 75~m long fiducial volume (FV) starting 2.6~m downstream of GTK3. The beam angular spread at the FV entrance is 0.11~mrad (rms) in both horizontal and vertical planes. Downstream of the FV, undecayed beam particles continue their path in vacuum.

Three-momenta of charged particles produced in $K^+$ decays are measured by a magnetic spectrometer (STRAW) located in the vacuum tank downstream of the FV. The spectrometer consists of four tracking chambers made of straw tubes, and a large aperture dipole magnet (M), located between the second and third chamber, that provides a horizontal momentum kick of 270~MeV/$c$. The momentum resolution is $\sigma_p/p = (0.30\oplus 0.005 \cdot p)\%$, with the momentum $p$ expressed in GeV/$c$.

A ring-imaging Cherenkov detector (RICH) consisting of a 17.5~m long vessel filled with neon at atmospheric pressure (with a Cherenkov threshold of 12.5~GeV/$c$ for pions) provides particle identification, charged particle time measurements (to a 70~ps accuracy for particles well above the Cherenkov threshold), and the trigger time. Two scintillator hodoscopes (CHOD), which include a matrix of tiles and two planes of slabs arranged in four quadrants located downstream of the RICH, provide trigger signals and time measurements. The tile matrix hodoscope has a time resolution of 1~ns, while the slab hodoscope measures time with 200~ps precision.

A $27X_0$ thick quasi-homogeneous liquid krypton (LKr) electromagnetic calorimeter is used for particle identification and photon detection. The calorimeter has an active volume of 7~m$^3$ segmented in the transverse direction into 13248 projective cells of $2\times2$~cm$^2$ size, and provides an energy resolution $\sigma_E/E=(4.8/\sqrt{E}\oplus11/E\oplus0.9)\%$, with $E$ expressed in GeV. To achieve hermetic acceptance for photons emitted in $K^+$ decays in the FV at angles up to 50~mrad from the beam axis, the LKr calorimeter is supplemented by annular lead glass detectors (LAV) installed in 12~positions inside and downstream of the vacuum tank, and two lead/scintillator sampling calorimeters (IRC, SAC) located close to the beam axis. An iron/scintillator sampling hadronic calorimeter formed of two modules (MUV1,2) and a muon detector consisting of 148~scintillator tiles located behind an 80~cm thick iron wall (MUV3) are used for particle identification. The~eight smaller tiles of MUV3 adjacent to the beam pipe are referred to as the \textit{inner tiles}, while the remaining 140 regular tiles are called the \textit{outer tiles}.

The data sample used for this analysis is obtained from $0.84\times 10^6$ SPS spills collected in 2017--2018, with the typical beam intensity increasing over time from \mbox{$1.5\times 10^{12}$} to \mbox{$2.2\times 10^{12}$} protons per spill. The latter value corresponds on average to a 500~MHz instantaneous beam particle rate at the FV entrance, and a 3.7~MHz $K^+$ decay rate in the FV. The main trigger of NA62 is dedicated to the collection of the very rare $K^+ \to \pi^+ \nu \bar{\nu}$ decays~\cite{PNN-2021}. Multi-track (MT) and di-muon multi-track ($2\mu$MT) triggers considered in this analysis operate concurrently, downscaled by typical factors of 100 and 2, respectively. The downscaling factors of both triggers were varied throughout the data taking. The MT trigger line selects the $\ktpil$~($\ktpis$) decays, used for normalization, while the $2\mu$MT line selects the $\kpimumus$ signal decays. The low-level hardware (L0) trigger~\cite{NA62-trigger} for both lines is based on RICH signal multiplicity and coincidence of signals in two opposite CHOD quadrants. The $2\mu$MT line additionally involves a requirement of signal coincidence in two outer MUV3 tiles. The high-level software (L1) trigger requires $K^+$ identification by KTAG, and reconstruction of a negatively charged STRAW track for both MT and $2\mu$MT trigger lines. A detailed description of the NA62 trigger system and its performance is given in~\cite{TriggerPerformance}.

Monte Carlo (MC) simulations of particle interactions with the detector and its response are performed using a software package based on the \geant toolkit~\cite{geant4}. In addition, the accidental activity is simulated, and the response of both trigger lines is emulated.

\section{Event selection}
\label{sec:selections}

Kinematic similarities of the signal ($\kpimumus$) and normalization ($\ktpis$) decays allow for substantial overlap between the signal and normalization event selections, which results in first-order cancellation of most detector and trigger inefficiencies, thus reducing the systematic uncertainties in the measurement.

~

\noindent The following selection criteria are common to the $\kpimumus$ and $\ktpis$ event selections.
\begin{itemize}
    \item Each STRAW track is assigned a time computed as a weighted average of the associated CHOD hodoscope signals. The weights are obtained from the time resolutions of the CHOD hodoscopes. Triplets of STRAW tracks compatible with a common origin in the FV are combined into three-track vertices. Vertex time is defined as the weighted average of the times of CHOD signals associated with the vertex tracks.
    \item Exactly one three-track vertex with the following properties is required to be present: total charge $q=1$, time within $6$~ns of the trigger time, $Z$ position between $110$~m and $180$~m from the target, total momentum compatible with the mean beam momentum within $2.5$~GeV$/c$, total transverse momentum with respect to the beam axis below $30$~MeV$/c$, and vertex distance from the beam axis below $5$~cm. The beam axis and momentum are monitored throughout the data taking with fully reconstructed $\ktpis$ decays. Only the three tracks forming the chosen vertex are considered in the following.
    \item All track times must be within $12$~ns of the vertex time, and the vertex time is required to be within $6$~ns of a KTAG kaon signal.
    \item The tracks must be within the geometrical acceptance of all STRAW chambers, and extrapolate to lie within the CHOD, LKr, and MUV3 acceptances.
    \item The track momenta should exceed $10$~GeV$/c$ to ensure track reconstruction efficiency above 90\%. The angles between each track and the beam axis must be smaller than $9$~mrad to reduce background to the $\kpimumus$ sample from $\ktpis$ decays followed by $\pi^\pm\to\mu^\pm\nu$ decays.
    \item The spatial separation between each pair of vertex tracks must be at least $15$~mm in the plane of the first STRAW chamber and $200$~mm in the LKr front plane to suppress photon conversions and the overlap of energy deposits.
\end{itemize}

\noindent The following particle identification criteria are employed.
\begin{itemize}
    \item A track is identified as a charged pion if it has no spatially associated MUV3 signals within $10$~ns of the vertex time, and the ratio of the associated LKr cluster energy to the track momentum is $E/p<0.9$.
    \item A track with $E/p<0.2$ is identified as a muon if it has a spatially associated MUV3 signal in an outer tile within $6$~ns of both the vertex and the trigger times.
\end{itemize}

\noindent The following criteria are specific to the $\kpimumus$ event selection.
\begin{itemize}
    \item Only vertices with tracks identified as $\pi^+\mu^+\mu^-$ are considered.
    \item To reduce the background from $\ktpis$ decays occurring upstream of the FV, the track identified as a $\pi^+$ is extrapolated backward to the COL plane, taking into account the TRIM5 magnetic field. The extrapolated position is required to lie outside a rectangle defined by $\vert X \vert < 40$~mm and $\vert Y \vert < 25$~mm.
    \item Further $\ktpis$ background suppression is achieved by requiring the momenta of both muon tracks to be below $45$~GeV$/c$.
    \item The invariant mass $m(\pi\mu\mu)$ of the three selected tracks is reconstructed with a $1.1$~MeV$/c^2$ resolution and must be within $8$~MeV$/c^2$ of the nominal $K^+$ mass $m_K$~\cite{PDG-2022}.
\end{itemize}

\noindent The following criteria are specific to the $\ktpis$ event selection.
\begin{itemize}
    \item In order to minimize differences between the signal and normalization selections, only one positive track, chosen at random, is required to be identified as a $\pi^+$.
    \item The identified $\pi^+$ track extrapolated to the COL plane must satisfy the same requirements as the $\pi^+$ in the $\kpimumus$ selection.
    \item The invariant mass $m(3\pi)$ of the three selected tracks is reconstructed with a $0.8$~MeV$/c^2$ resolution and must be within $8$~MeV$/c^2$ of $m_K$.
\end{itemize}

\noindent For both selections, simulated events are required to be accepted by a set of software algorithms emulating the conditions employed in the online trigger system.

\section{Signal and normalization samples}
\label{sec:sig_norm}

\begin{figure}[htb]
    \centering
    \includegraphics[width=0.49\columnwidth, page=1, trim={0cm 0.8cm 0cm 0.9cm}, clip]{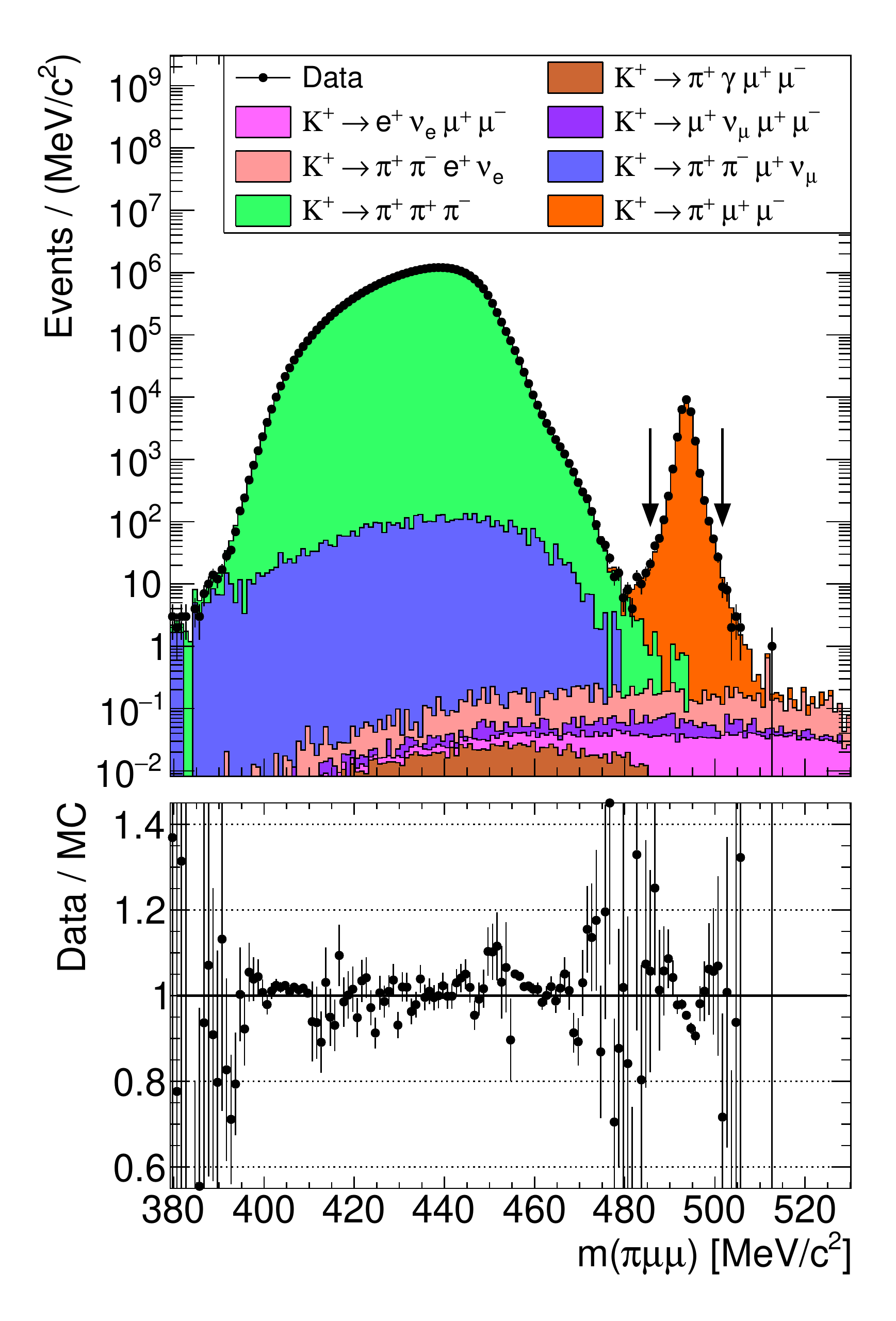}
    \includegraphics[width=0.49\columnwidth, page=1, trim={0cm 0.8cm 0cm 0.9cm}, clip]{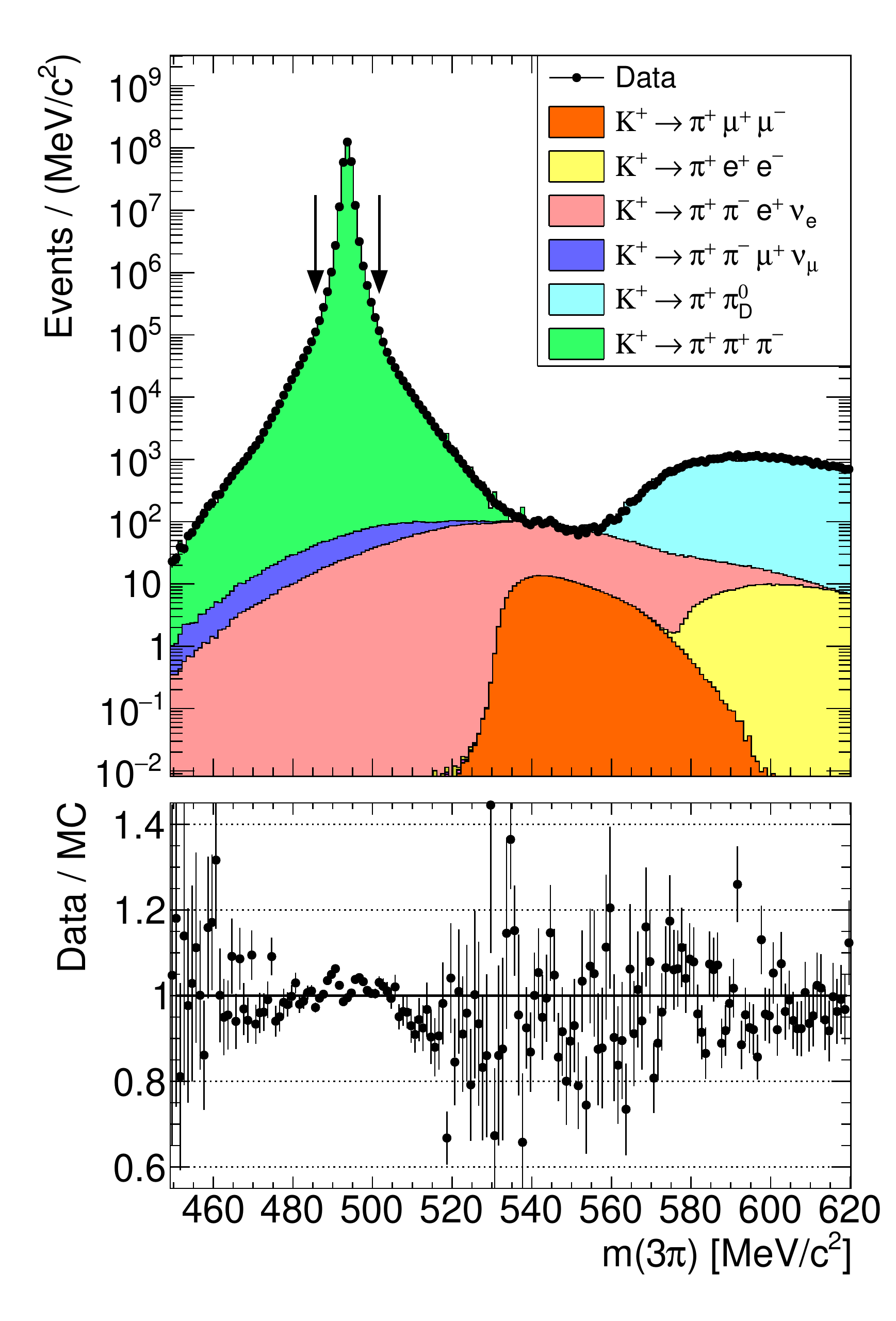}
    
    \caption{\textbf{Top:} reconstructed mass distributions of events satisfying the signal (left) and normalization (right) selections. The arrows indicate the selected mass regions. The contribution from the simulated $\kpimumus$ decays is scaled according to the PDG branching fraction~\cite{PDG-2022}. \textbf{Bottom:} ratios of data and simulated spectra for signal (left) and normalization (right).}
    \label{fig:stack_minv}
\end{figure}

The reconstructed mass spectra of the data and simulated events passing the signal and normalization event selections are shown in \autoref{fig:stack_minv}. The selected $\ktpis$ data sample, contaminated by background decays to a negligible level of $10^{-6}$, is used together with the simulated $\ktpis$ events with inner bremsstrahlung included~\cite{K3pig}, to obtain the effective number of kaon decays in the FV
\begin{equation}
    N_K = \frac{1}{A_{3\pi}\cdot\mathcal{B}_{3\pi}} \cdot \mathlarger{\mathlarger{\sum_{i}} ~} \frac{N^i_{3\pi} \cdot D^i_\text{MT}}{D^i_{2\mu\text{MT}}} = (3.48 \pm 0.09_\text{syst} \pm 0.02_\text{ext})\times 10^{12},
    \label{eq:NK}
\end{equation}
where the index $i$ runs over data taking periods defined by constant trigger downscaling factors, $N_{3\pi}^i$ are the numbers of $\ktpis$ events selected with the MT trigger with downscaling factor $D^i_\text{MT}$, $D^i_{2\mu\text{MT}}$ are the downscaling factors of the $2\mu$MT trigger, and $A_{3\pi} = (6.58 \pm 0.16)\%$  and $\mathcal{B}_{3\pi} = (5.583 \pm 0.024)\%$ are the acceptance (obtained from simulation) and the branching fraction~\cite{PDG-2022} of the $\ktpis$ decay, respectively. The statistical errors in $A_{3\pi}$ and $N_K$ are negligible, while the systematic uncertainties are dominated by the accuracy of the CHOD detector efficiency in the simulation. The external error on $N_K$ stems from the uncertainty on the $\ktpis$ branching fraction.

The $m(\pi\mu\mu)$ signal region contains $27679$ data events with a background contamination of about 8 events, estimated from simulation.

\section{Interpretation of the data}
\label{sec:results}

\subsection{Decay width and form factor parameterization}

The one-photon-inclusive $\kpimumus$ differential decay width expressed in terms of the normalized $\mu^+\mu^-$ invariant mass squared $z = m^2(\mu^+ \mu^-)/m_K^2$ reads~\cite{D'Ambrosio:1998, Dubnickova:2006, Kubis:2010, TomasKpmm}
\begin{equation}
\label{eq:dGdzdx}
\frac{\text{d}\Gamma(z)}{\text{d}z} = \frac{\text{d}\Gamma_\text{3-body}(z)}{\text{d}z} + \frac{\text{d}\Gamma_\text{4-body}(z)}{\text{d}z} = g(z) 
\cdot \vert W(z) \vert^2 + \frac{\text{d}\Gamma_\text{4-body}(z)}{\text{d}z},
\end{equation}
where $(2 m_\mu / m_K)^2<z<(1-m_\pi / m_K)^2$, $W(z)$ is the form factor of the $K^+ \to \pi^+ \mu^+ \mu^-$ transition, and $g(z)$ is a function describing the decay kinematics~\cite{Dubnickova:2006} and including next-to-leading order electromagnetic effects in terms of radiative corrections. While the $\mu^+\mu^-$ interactions are fully taken into account by virtual and bremsstrahlung corrections for the lepton and meson contributions, discussed in~\cite{Kubis:2010} and extended beyond the soft-photon approximation, the semi-classical Coulomb corrections, summarized for example in~\cite{Coulomb}, are applied to the $\pi^+\mu^+$ and $\pi^+\mu^-$ pairs. These last corrections have opposite sign and the same average magnitude; their combined effect on the results of the present analysis is found to be negligible. The hard-photon 4\nobreakdash-body ($K^+ \to \pi^+\mu^+\mu^- \gamma$) part of the phase-space is separated from the soft-photon 3\nobreakdash-body ($K^+ \to \pi^+\mu^+\mu^-$) part by the condition $(P_{\pi} + P_\gamma)^2 - m_{\pi}^2 > 100$~MeV$^2$, where $P_{\pi}$ and $P_\gamma$ are 4-momenta of the $\pi^+$ and~$\gamma$, respectively. The cutoff value is optimized with respect to the experimental resolution. The resulting ratio of the 4\nobreakdash-body to 3\nobreakdash-body integrated decay widths is $(1.64 \pm 0.02)\%$, where the uncertainty comes mainly from the accuracy of the theoretical description $\text{d}\Gamma_\text{4-body}(z)/\text{d}z$ of the 4\nobreakdash-body decay~\cite{TomasKpmm}. In the present analysis, the 4\nobreakdash-body decay width, depending non-trivially on the form factor, is approximated by a unique function displayed in \autoref{fig:dGdz_W2}-left. Effects of this approximation are treated as systematic uncertainties.

The Chiral Perturbation Theory parameterization of $W(z)$ at $\mathcal{O}(p^6)$, introduced in~\cite{D'Ambrosio:1998}, is used in the present paper:
\begin{equation} 
W(z) = G_F m_K^2 (a_+ + b_+ z) + W^{\pi\pi}(z),
\label{eq:FF}
\end{equation}
\noindent where $a_+$ and $b_+$ are real parameters, and $W^{\pi\pi}(z)$ is a complex function describing the contribution from a two-pion loop. The term $W^{\pi\pi}(z)$ depends on additional real parameters $\alpha_+$ and $\beta_+$; the values $\alpha_+ = (-20.40 \pm 0.18)\times10^{-8}$ and $\beta_+ = (-2.05 \pm 0.06)\times10^{-8}$~\cite{Knecht-newBeta} are used.

\subsection{Measurement of the model-independent branching fraction and form factor}

The selected $\kpimumus$ signal sample with negligible background contamination is distributed in 50 equipopulated bins in $z$ with widths ranging from $0.004$ for $z \approx 0.25$ to $0.066$ for the last bin. The resolution in $z$ increases linearly from zero to $0.0035$ within the allowed kinematic range, and is always several times smaller than the corresponding bin width.

\begin{figure}[t]
    \centering
    \includegraphics[width=0.49\columnwidth, page=1, trim={0cm 0cm 0cm 0.0cm}, clip]{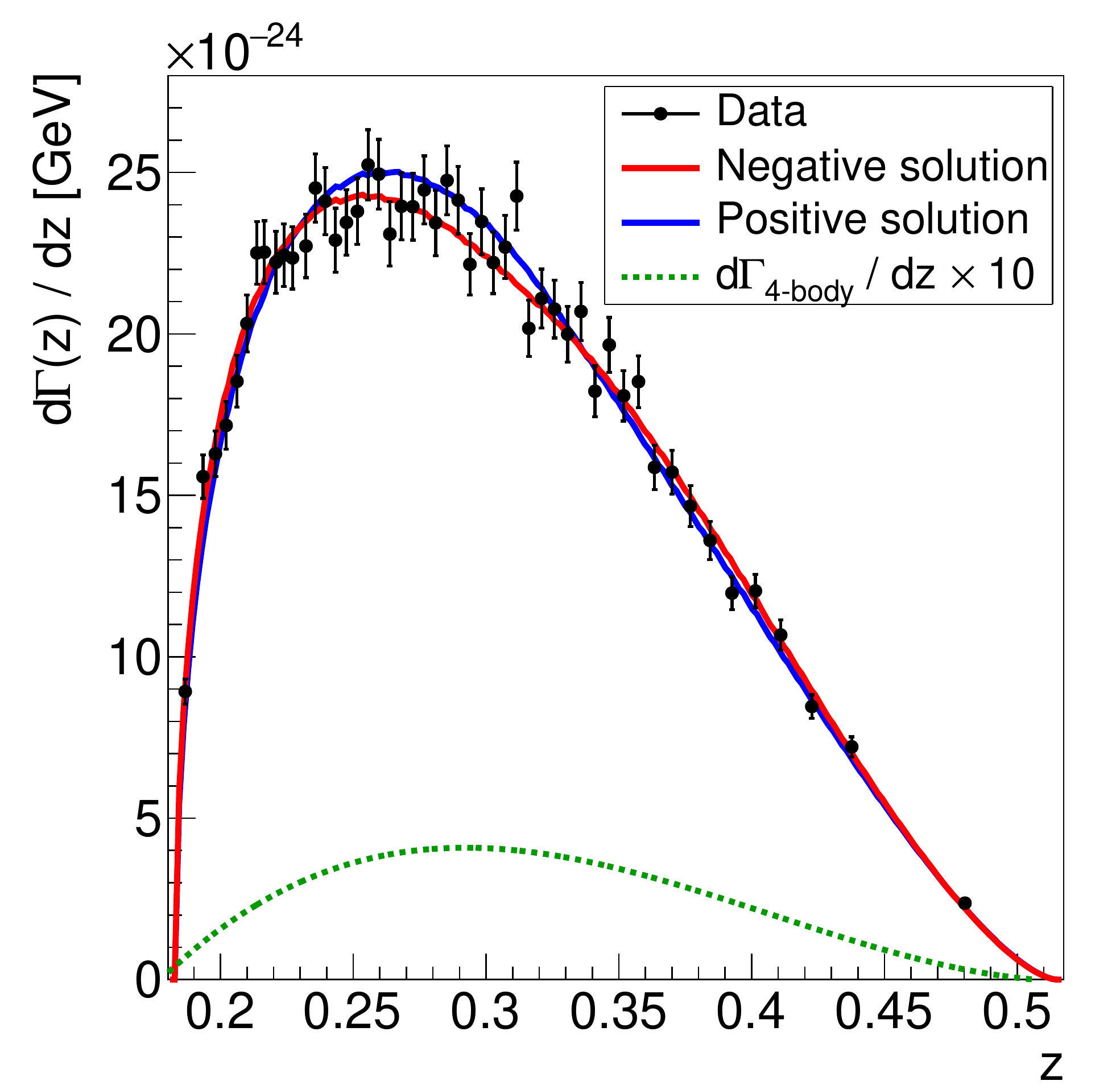}
    \includegraphics[width=0.49\columnwidth, page=1, trim={0cm 0cm 0cm 0.0cm}, clip]{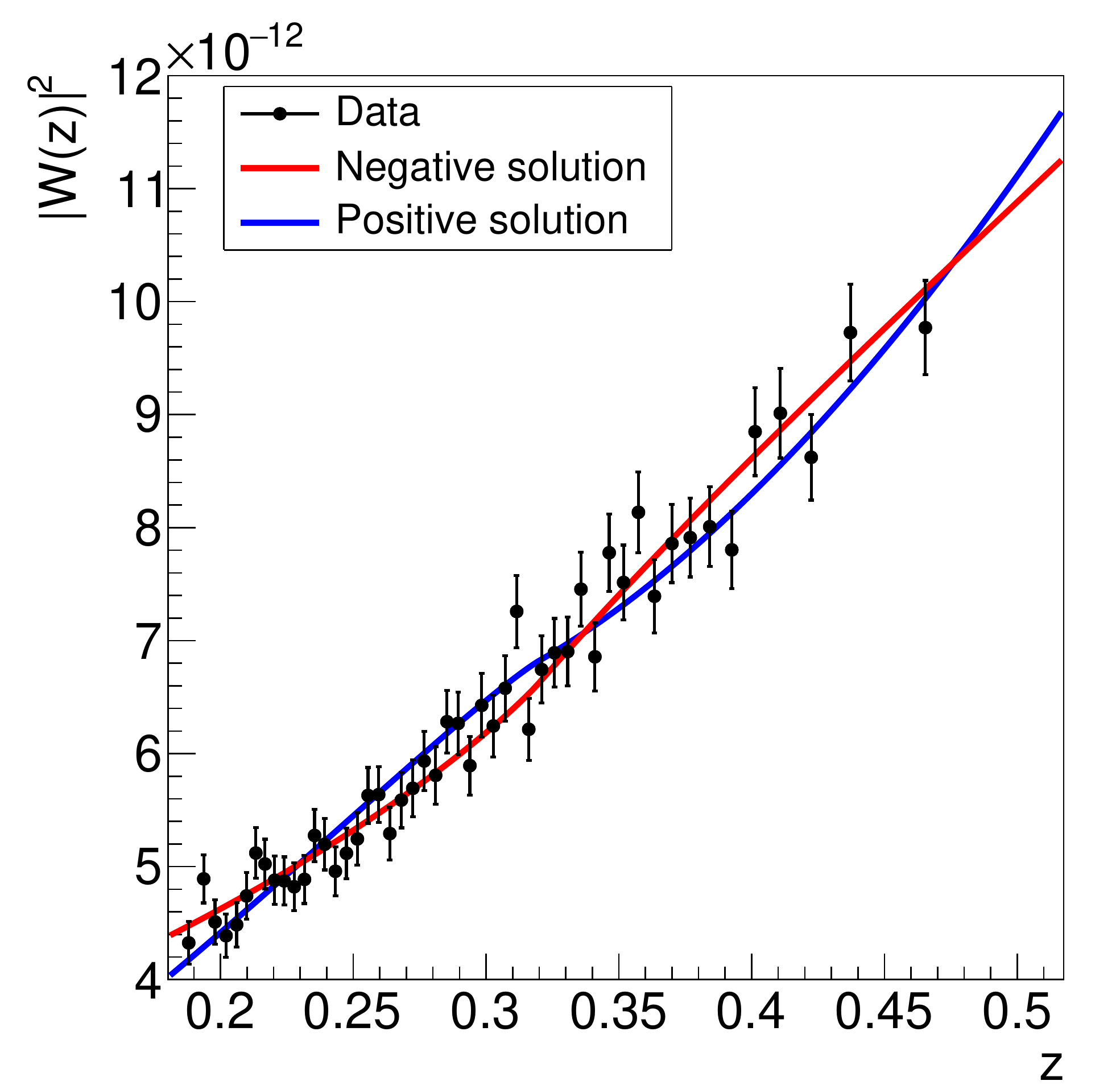}
    
    \caption{\textbf{Left:} reconstructed $\kpimumus$ differential decay width together with the 4-body decay contribution scaled by a factor of 10. The horizontal positions of the data points take into account the non-linearity of the fit function~\cite{WhereToStick}. \textbf{Right:} squared modulus of the form factor. The superimposed lines correspond to the two form factor fit solutions defined in the text.}
    \label{fig:dGdz_W2}
\end{figure}

\begin{figure}[t]
    \centering
    \includegraphics[width=0.49\columnwidth, page=1, trim={0.5cm 0.5cm 0cm 0.9cm}, clip]{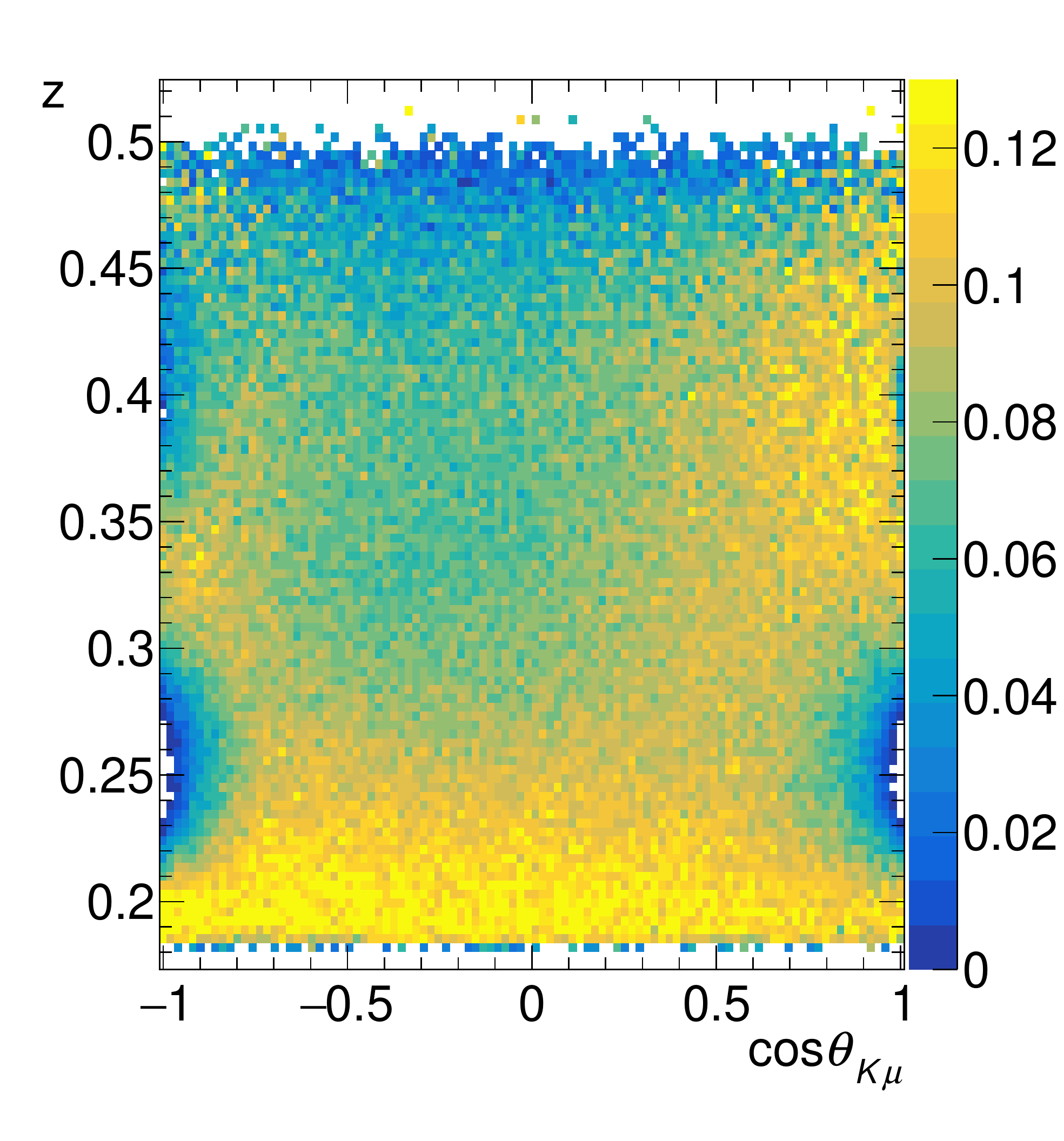}
    \hspace{+0.2cm}\includegraphics[width=0.49\columnwidth, page=1, trim={0cm 0.3cm 0.3cm 0.9cm}, clip]{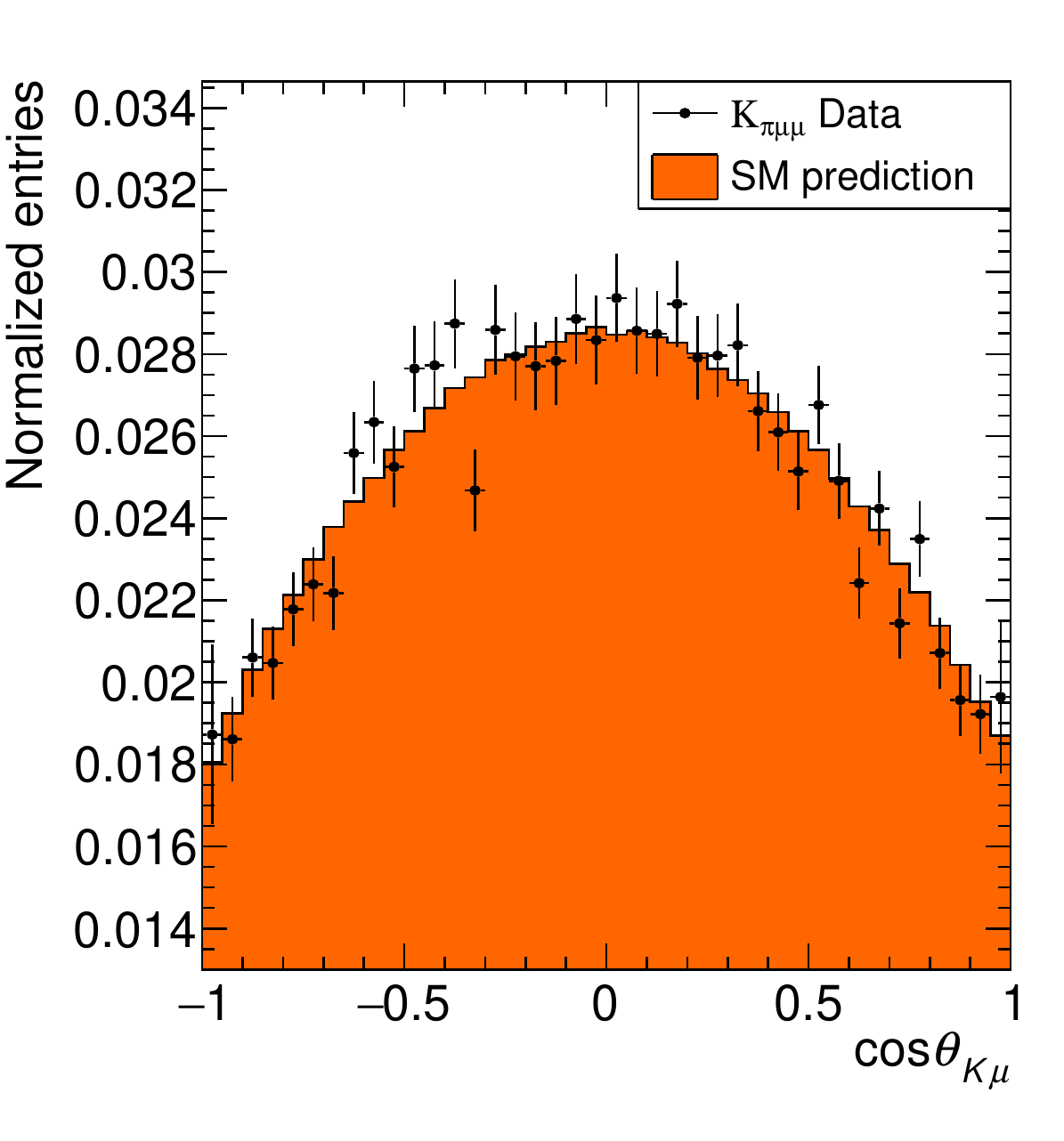}

    \vspace{-0.4cm}
    \caption{\textbf{Left:} signal selection acceptance of the simulated $\kpimumus$ decays in the ($\cos\theta_{K\mu}$,~$z$) plane. \textbf{Right:} reconstructed $\cos\theta_{K\mu}$ spectrum from data corrected for the acceptance (points with error bars), and the distribution expected from the Standard Model (filled histogram).}
    \label{fig:COSspectrum}
\end{figure}

The reconstructed differential decay width, shown in \autoref{fig:dGdz_W2}-left, is given by
\begin{equation}
    \left(\frac{\text{d}\Gamma(z)}{\text{d}z}\right)_i = \frac{N_{\pi\mu\mu, i}}{A_{\pi\mu\mu, i}} \cdot \frac{1}{\Delta z_i} \cdot \frac{1}{N_K} \cdot \frac{\hbar}{\tau_K},
    \label{eq:binneddGdz}
\end{equation}
where for each bin $i$: $N_{\pi\mu\mu, i}$ is the number of $\kpimumus$ signal candidates, $\Delta z_i$ is the bin width, $A_{\pi\mu\mu, i}$ is the signal selection acceptance of the $\kpimumus$ decay (obtained from simulation, and equal to zero at both kinematic bounds of $z$ while reaching the maximum of 12.5\% around $z = 0.2$, see also \autoref{fig:COSspectrum}-left), $N_K$ is the effective number of kaon decays in the FV collected by the $2\mu$MT trigger (\autoref{eq:NK}), $\hbar$ is the reduced Planck constant, and $\tau_K = (1.238 \pm 0.002)\times 10^{-8}$~s is the mean charged kaon lifetime~\cite{PDG-2022}.

\newpage

The model-independent $\kpimumus$ branching fraction
\begin{equation*}
\mathcal{B}_{\pimumug} = (9.15 \pm 0.06_\text{stat})\times 10^{-8}
\end{equation*}
is obtained from the reconstructed binned differential decay width (\autoref{eq:binneddGdz}, \autoref{fig:dGdz_W2}-left) by integrating the spectrum over $z$ and multiplying by $\tau_K/\hbar$.

The $\kpimumus$ data sample is also used to extract the $\vert W(z) \vert^2$ form factor (\mbox{\autoref{fig:dGdz_W2}-right}). The values of the $\vert W(z) \vert^2$ function are reconstructed from the differential decay spectrum (\autoref{fig:dGdz_W2}-left) under the assumption that $\vert W(z) \vert^2$ is linear in each bin of~$z$. This assumption defines the horizontal positions of the data points in \autoref{fig:dGdz_W2}-right, which are different from the positions in \autoref{fig:dGdz_W2}-left.

The form factor parameters $a_+$ and $b_+$ best describing the data are determined by a $\chi^2$ fit of the data points shown in \autoref{fig:dGdz_W2}. Fits of $\text{d}\Gamma(z)/\text{d}z$ and $\vert W(z) \vert^2$ give identical results. The theoretically-preferred~\cite{FitOfBeta} \textit{negative solution} with both $a_+$ and $b_+$ negative and $\chi^2/\text{ndf} = 45.1/48$ ($p\text{-value} = 0.59$) is

\vspace{-6mm}
\begin{align*}
    a_+ = -0.575 \pm 0.012_\text{stat}, ~~ b_+ = -0.722 \pm 0.040_\text{stat}, ~~ \text{with correlation } \rho(a_+, b_+) = -0.972.
\end{align*}
A~second $\chi^2(a_+, b_+)$ minimum is found, corresponding to the \textit{positive solution}: $\chi^2/\text{ndf} = 56.4/48$ ($p\text{-value} = 0.19$), $a_+ = 0.373 \pm 0.012_\text{stat}$, $b_+ = 2.017 \pm 0.040_\text{stat}$, $\rho(a_+, b_+) = -0.973$. Only the negative solution is considered in the following.

\subsection{Forward-backward asymmetry measurement}

The forward-backward asymmetry $A_\text{FB}$ of the $\kpimumus$ decay is defined in terms of the angle $\theta_{K\mu}$ between the $K^+$ and the $\mu^-$ three-momenta in the $\mu^+\mu^-$ rest frame, as
\begin{equation}
     A_\text{FB} = \frac{\mathcal{N}(\cos\theta_{K\mu}>0) - \mathcal{N}(\cos\theta_{K\mu}<0)}{\mathcal{N}(\cos\theta_{K\mu}>0)+\mathcal{N}(\cos\theta_{K\mu}<0)},
\end{equation}
\noindent where the numbers of events $\mathcal{N}$ are obtained after correction for the non-uniform acceptance in the ($\cos\theta_{K\mu}$, $z$) plane (\autoref{fig:COSspectrum}-left). The resulting $\cos\theta_{K\mu}$ spectrum of the data events and the distribution expected from the Standard Model (SM) are displayed in \autoref{fig:COSspectrum}-right.

The asymmetry is measured to be
\begin{equation*}
    A_\text{FB} = (0.0 \pm 0.7_\text{stat}) \times 10^{-2}
\end{equation*}
and shows no significant dependence on $z$. The statistical precision is at the level of the upper limits on $A_\text{FB}$ predicted by the Minimal Supersymmetric Standard Model~\cite{FBasymmetry-MSSM} and by the calculation of the two-photon intermediate state $K^+ \to \pi^+ \gamma^* \gamma^* \to \pi^+ \mu^+ \mu^-$~\cite{FBasymmetry-2gamma}.

\section{Systematic and external uncertainties}
\label{sec:systematics}

The individual contributions to the total uncertainties are discussed in the following and listed in \autoref{tab:error_budget}.

\subsection{Trigger efficiency}

The trigger behaviour is emulated with a set of software algorithms applied to simulated events. The algorithms are tuned and validated on $\ktpis$ events. The L0 RICH, L0 CHOD, L0 MUV3 and L1 KTAG trigger efficiencies (equal to 99.8\%, 98.2\%, 98.9\% and 99.8\%, respectively) are found to be independent of the decay kinematics. Data and simulation efficiencies agree within $0.3\%$. The L1 STRAW trigger efficiency is 94.7\% and varies as a function of decay kinematics within $\mathcal{O}(1\%)$. Data and simulation efficiencies agree within $0.5\%$.

The similarity of the MT and $2\mu$MT trigger lines results in substantial cancellation of trigger-related systematic effects. The residual systematic uncertainties are estimated by either disabling the software trigger emulators in simulation (in the case of the L0 RICH and L0 CHOD conditions), or replacing them with simplified emulators (L0 MUV3, L1 KTAG, L1 STRAW).

\subsection{Reconstruction and particle identification}

The similarity of the signal and normalization selections allows for significant cancellation of most systematic effects coming from reconstruction and particle identification efficiencies.

Systematic uncertainties arising from differences between event reconstruction efficiencies in data and simulation are dominated by the three-track event reconstruction in the STRAW spectrometer. A dedicated $K_{3\pi}$ event selection, relying on a reconstructed kaon track in the GTK and two pion tracks in the STRAW, is used to measure the efficiency of reconstructing the third pion track. The average measured efficiency is $84\%$ and depends on the decay kinematics. The~observed differences of up to $2\%$ between the efficiencies in data and simulation are considered in evaluating the systematic effects resulting from the STRAW track reconstruction efficiency.

The CHOD and MUV3 reconstruction efficiencies are above 99\%, with no more than 0.6\% difference between data and simulation.

The differences between data and simulation in the hadronic shower development and energy reconstruction in the LKr are another source of systematic uncertainty. No significant difference is observed in the efficiency of the muon identification. The efficiency of the pion identification measured on data is 99\%. The agreement between data and simulation varies with pion momentum within 1\%. Residual effects due to different $\kpimumus$ and $\ktpis$ pion kinematics are treated as systematic uncertainties.

\subsection{Beam and accidental activity simulation}

Systematic uncertainties stemming from the quality of the simulation of the beam momentum spectrum and intensity profile, and from the accuracy of the simulation of the halo muons accompanying the beam, are combined into a single systematic uncertainty. The selected normalization sample of $\ktpis$ events is used for the beam momentum and intensity studies. The halo muons are selected from out-of-time STRAW tracks that have associated signals in MUV3 and are not compatible with originating from decays in the FV.

\subsection{Background}

The number of background events is estimated using simulation to be $7.8 \pm 5.6$, where the error comes from the limited statistics of simulated background decays. The background arises mainly from the $\ktpis$ contribution with two $\pi^\pm\to\mu^\pm\nu$ decays in flight. More details on the methods employed in the $\ktpis$ background estimation can be found in~\cite{LNV-K3piBackground}.

Systematic uncertainties from the background contamination are estimated conservatively as differences between the results obtained with background neglected and background subtracted.

\subsection{External uncertainties}

External uncertainties in the measured quantities originate from the $\ktpis$ branching fraction~\cite{PDG-2022}, from the accuracy of the radiative corrections to the $\kpimumus$ decay, including the numerical approximation of $\text{d}\Gamma_\text{4-body}/\text{d}z$, and from the pion loop term parameters $\alpha_+$ and $\beta_+$~\cite{Knecht-newBeta}.

\begin{table}[ht]
    \caption{Summary of uncertainties.}
    \vspace{2mm}
    \centering
    \begin{tabular}{l|c|c|c|c}
     & $\delta a_+$ & $\delta b_+$ & $\delta \mathcal{B}_{\pimumug} \times 10^{8}$ & $\delta A_\text{FB}\times 10^{2}$  \\
    \toprule
    \textbf{Statistical uncertainty}         & $0.012$ & $0.040$ & $0.06$ & $0.7$  \\
    \bottomrule
    \toprule
    Trigger efficiency         & $0.002$ & $0.008$ & $0.02$ & $0.1$  \\
    Reconstruction and particle identification  & $0.002$ & $0.007$ & $0.02$ & $0.1$  \\
    Size of the simulated $\kpimumus$ sample & $0.002$ & $0.007$ & $0.01$ & $0.1$  \\
    Beam and accidental activity simulation & $0.001$ & $0.002$ & $0.01$ & ---  \\
    Background                 & $0.001$ & $0.001$ & ---    & ---     \\
    \midrule
    \textbf{Total systematic uncertainty}         & $0.003$ & $0.013$ & $0.03$ & $0.2$  \\
    \bottomrule
    \toprule
    $\ktpis$ branching fraction                             & $0.001$ & $0.003$ & $0.04$ & ---     \\
    $\kpimumus$ radiative corrections & $0.003$ & $0.009$ & $0.01$ & $0.2$  \\
    Parameters $\alpha_+$ and $\beta_+$                 & $0.001$ & $0.006$ & --- & ---     \\
    \midrule
    \textbf{Total external uncertainty}                                                   & $0.003$ & $0.011$ & $0.04$ & $0.2$  \\
    \bottomrule
    \toprule
    \textbf{Total uncertainty}                                                   & $\mathbf{0.013}$ & $\mathbf{0.043}$ & $\mathbf{0.08}$ & $\mathbf{0.7}$  \\
    \bottomrule
    \end{tabular}
    \label{tab:error_budget}
\end{table}

\vspace{-4mm}
\section{Comparison with earlier measurements}
\label{sec:world}

\begin{figure}[t]
  \centering
  \includegraphics[width=0.495\columnwidth, page=1, trim={0cm 0.0cm 0cm 0.0cm}, clip]{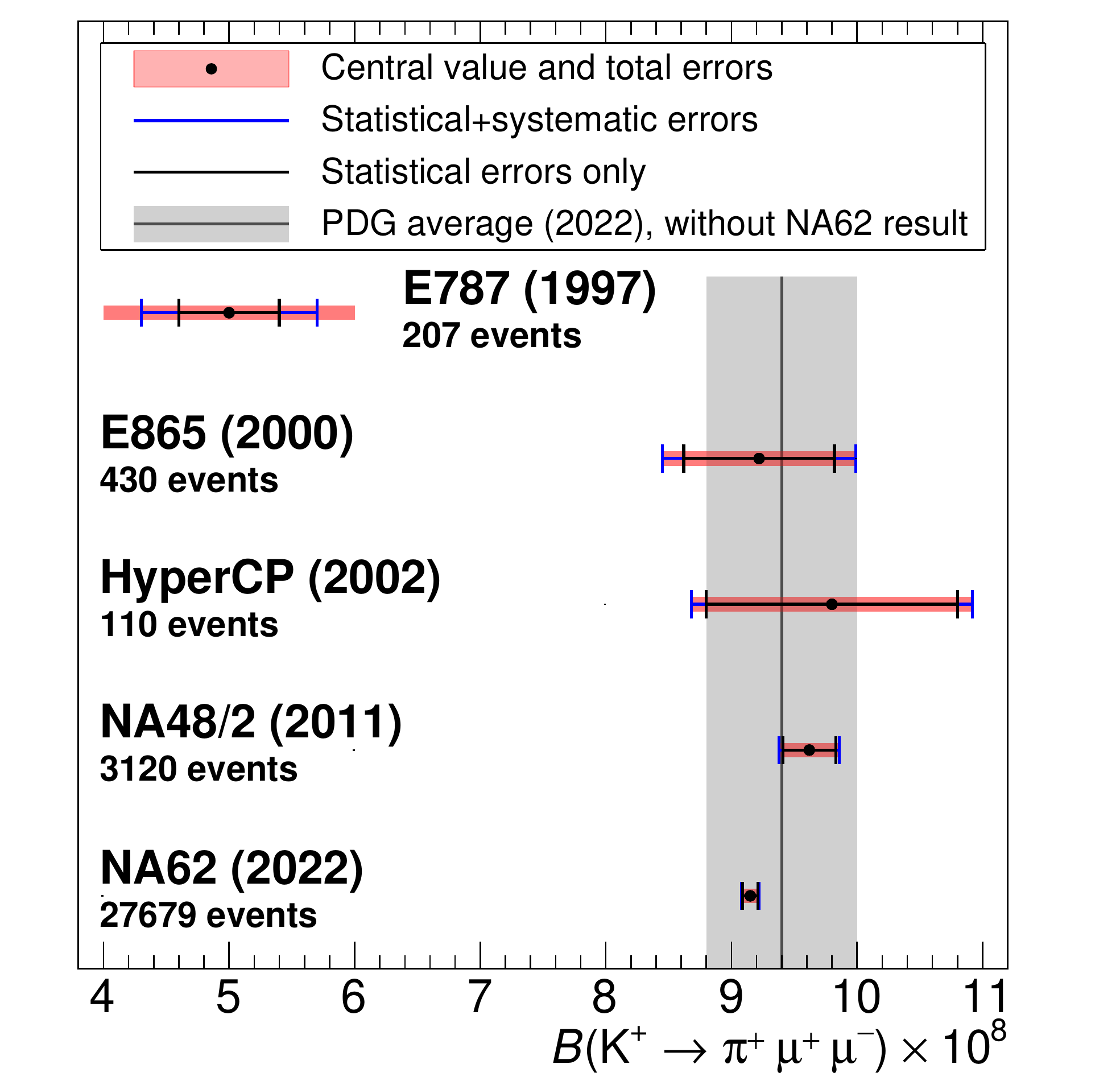}
  \includegraphics[width=0.495\columnwidth, page=1, trim={0cm 0.0cm 0cm 0.0cm}, clip]{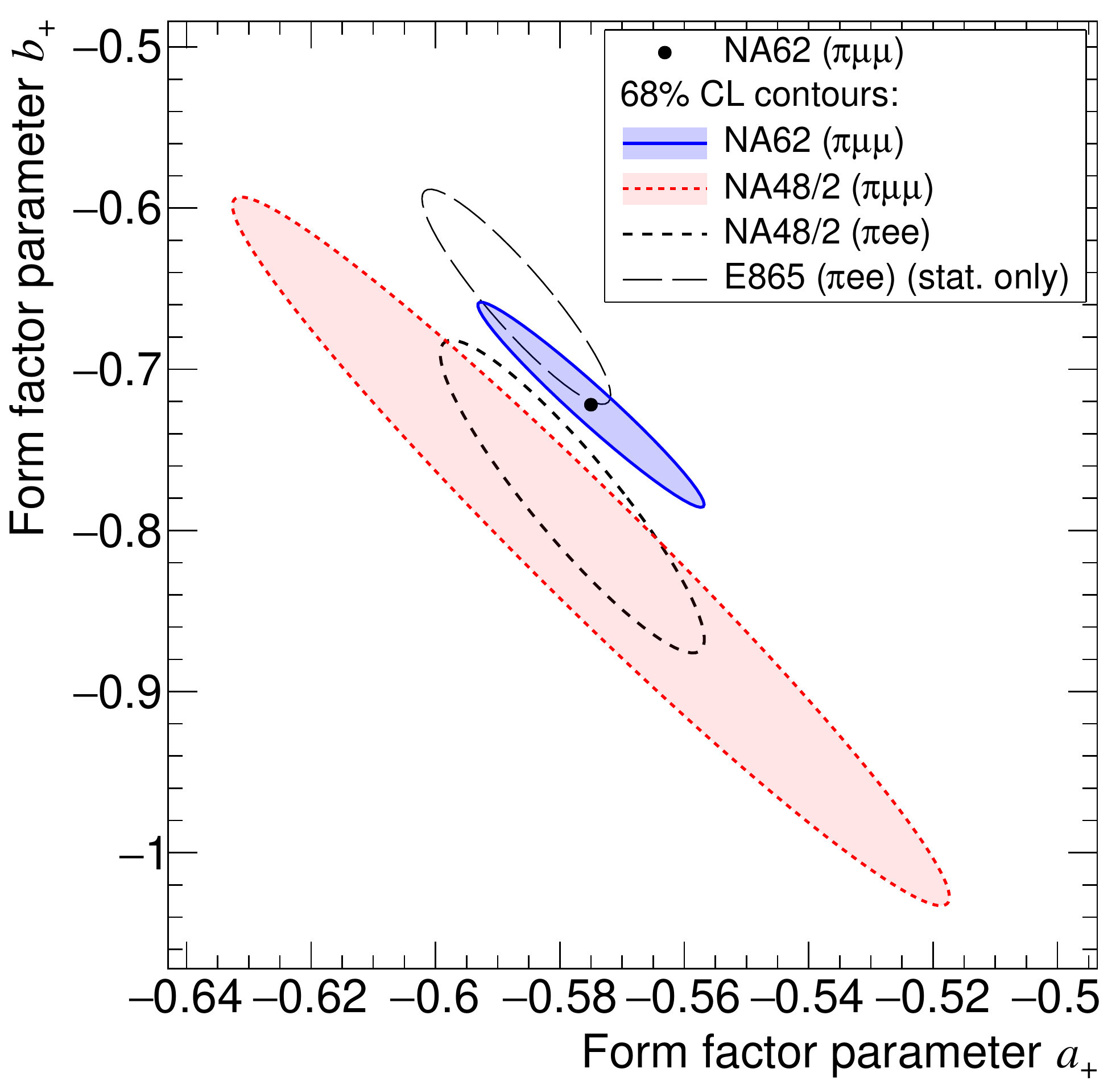}
   
  \vspace{-2mm}
  \caption{Comparison with earlier measurements. \textbf{Left:}~the $\kpimumus$ branching fraction, with the PDG~\cite{PDG-2022} average shown as a shaded band. \textbf{Right:}~combined statistical and systematic $68\%$ CL contours in the $(a_+, b_+)$ plane for the muon and electron modes. The NA48/2 value of $\rho(a_+, b_+)$ for the electron mode is used to construct the E865 contour, as E865~\cite{E865-piee} does not quote the correlation. The systematic uncertainties in $a_+$ and $b_+$ are not provided by E865~\cite{E865-piee}.}
  \label{fig:world}
\end{figure}

A comparison of the present results with those from previous measurements by E787, E865, HyperCP and NA48/2 is shown in \autoref{fig:world}, \autoref{tab:previousBR}, and \autoref{tab:previousFF}.

Note that the NA48/2 measurement~\cite{NA48/2-pimumu}, until now the most precise, used a different $\ktpis$ branching fraction~\cite{PDG-2010}, and did not simulate the inner bremsstrahlung radiation of $K_{3\pi}$ decays. Implementing these conditions in the NA62 analysis has minor impact on the results, which would change by $\delta a_+ = -0.001$, $\delta b_+ = -0.002$, and $\delta \mathcal{B}_{\pi\mu\mu} = +0.03\times 10^{-8}$.

Furthermore, the analysis by NA48/2 did not simulate inclusive radiative corrections and the 4-body radiative decay $K^+ \to \pi^+\mu^+\mu^-\gamma$ in the $\kpimumus$ sample, but implemented only the soft-photon Coulomb corrections for all pairs of the $K_{\pi\mu\mu}$ decay products. Adopting this approach changes the NA62 results by $\delta a_+ = -0.006$, $\delta b_+ = +0.034$, and $\delta \mathcal{B}_{\pi\mu\mu} = -0.06\times 10^{-8}$, where the $0.7\%$ relative change in the branching fraction comes from the increase of the signal acceptance measured with the 3-body simulated $K_{\pi\mu\mu}$ sample including Coulomb corrections.

In addition, previous experiments employed values of $\alpha_+ = -20.6\times10^{-8}$ and $\beta_+ = -2.8\times10^{-8}$, taken from~\cite{D'Ambrosio:1998}. Using these values instead of the revised ones ($\alpha_+ = -20.40\times10^{-8}$, $\beta_+ = -2.05\times10^{-8}$~\cite{Knecht-newBeta}), the NA62 results would change\footnote{The measured slopes are $\delta a_+/\delta \alpha_+=+0.004\times 10^8$, $\delta b_+/\delta \alpha_+=-0.029\times 10^8$, $\delta a_+/\delta \beta_+=+0.013\times 10^8$, and $\delta b_+/\delta \beta_+=-0.027\times 10^8$.} by $\delta a_+=-0.011$, $\delta b_+=+0.026$.

\vspace{-2mm}
\begin{table}[H]
    \caption{Comparison with the previous measurements of the $\kpimumus$ branching fraction.}
    \vspace{2mm}
    \centering
    \begin{tabular}{l|c|c}
    Measurement & $\kpimumus$ candidates & $\mathcal{B}_{\pimumug} \times 10^{8}$ \\
    \toprule
    E787~\cite{E787-pmm} & \phantom{00}$207$ & $5.0\phantom{0} \pm 0.4_\text{stat}\phantom{0} \pm 0.7_\text{syst}\phantom{0} \pm 0.6_\text{ext}\phantom{0} = 5.0\phantom{0} \pm 1.0\phantom{0}$ \\
    E865~\cite{E865-pmm} & \phantom{00}$430$ & $9.22 \pm 0.60_\text{stat} \pm 0.49_\text{syst} \phantom{ {}\pm{} 0.00_\text{ext}} = 9.22 \pm 0.77$ \\
    HyperCP~\cite{HyperCP} & \phantom{00}$110$ & $9.8\phantom{0} \pm 1.0_\text{stat}\phantom{0} \pm 0.5_\text{syst}\phantom{0 \pm 0.00_\text{ext}} = 9.8\phantom{0} \pm 1.1\phantom{0}$ \\
    NA48/2~\cite{NA48/2-pimumu} & \phantom{0}$3120$ & $9.62 \pm 0.21_\text{stat} \pm 0.11_\text{syst} \pm 0.07_\text{ext} = 9.62 \pm 0.25$ \\
    NA62, this result & $27679$ & $9.15 \pm 0.06_\text{stat} \pm 0.03_\text{syst} \pm 0.04_\text{ext} = 9.15 \pm 0.08$ \\
    \bottomrule
    \end{tabular}
    \label{tab:previousBR}
\end{table}

\vspace{-6mm}
\begin{table}[H]
    \caption{Comparison with the previous measurements of the $\kpillc$ form factor parameters. The E865 $\kpiees$ measurement~\cite{E865-piee} does not provide the systematic uncertainties, nor the correlation coefficient of the form factor parameters.}
    \vspace{2mm}
    \centering
    \begin{tabular}{l|c|c|c|c}
    Measurement & Signal candidates & $a_+$ & $b_+$ & $\rho(a_+, b_+)$ \\
    \toprule
    E865, $\kpiees$~\cite{E865-piee} & $10300$ & $-0.587 \pm 0.010$ & $-0.655 \pm 0.044$ & --- \\
    NA48/2, $\kpiees$~\cite{NA48/2-piee} & \phantom{0}$7253$ & $-0.578 \pm 0.016$ & $-0.779 \pm 0.066$ & $-0.913$ \\
    NA48/2, $\kpimumus$~\cite{NA48/2-pimumu} & \phantom{0}$3120$ & $-0.575 \pm 0.039$ & $-0.813 \pm 0.145$ & $-0.976$ \\
    NA62, $\kpimumus$, this result & $27679$ & $-0.575 \pm 0.013$ & $-0.722 \pm 0.043$ & $-0.972$ \\
    \bottomrule
    \end{tabular}
    
    \label{tab:previousFF}
\end{table}

\vspace{-3mm}
\section{Summary}
\label{sec:summary}

A sample of $27679$ $\kpimumus$ candidates with negligible background contamination was collected by the NA62 experiment in 2017--2018. The size of the $\kpimumus$ data sample is the main factor limiting the precision of the present analysis.

The $\kpimumus$ model-independent branching fraction is measured to be $(9.15 \pm 0.08)\times 10^{-8}$, consistent with previous measurements and at least a factor of three more precise.

The form factor parameters in the framework of the Chiral Perturbation Theory at $\mathcal{O}(p^6)$ are measured as $a_+ = -0.575 \pm 0.013$, $b_+ = -0.722 \pm 0.043$. Values and statistical errors of parameters in any other form factor model can be obtained from the reconstructed values of the $\vert W(z) \vert^2$ function. The present measurement is the first to employ inclusive radiative corrections in the simulation of the signal channel. The form factor parameters are consistent with those measured by NA48/2, as well as with the results obtained in the electron mode by other experiments, suggesting agreement with lepton flavour universality in the $\kpills$ decays.

The forward-backward asymmetry of the $\kpimumus$ decay is measured to be $A_\text{FB} = (0.0 \pm 0.7)\times 10^{-2}$, a factor of $2.6$ improvement in the precision with respect to NA48/2. The experimental precision reaches the level of the upper limits on $A_\text{FB}$ predicted by the Minimal Supersymmetric Standard Model and by the calculation of the two-photon intermediate state $K^+ \to \pi^+ \gamma^* \gamma^* \to \pi^+ \mu^+ \mu^-$.

\section*{Acknowledgements}
It is a pleasure to express our appreciation to the staff of the CERN laboratory and the technical
staff of the participating laboratories and universities for their efforts in the operation of the
experiment and data processing.

The cost of the experiment and its auxiliary systems was supported by the funding agencies of 
the Collaboration Institutes. We are particularly indebted to: 
F.R.S.-FNRS (Fonds de la Recherche Scientifique - FNRS), under Grants No. 4.4512.10, 1.B.258.20, Belgium;
CECI (Consortium des Equipements de Calcul Intensif), funded by the Fonds de la Recherche Scientifique de Belgique (F.R.S.-FNRS) under Grant No. 2.5020.11 and by the Walloon Region, Belgium;
NSERC (Natural Sciences and Engineering Research Council), funding SAPPJ-2018-0017,  Canada;
MEYS (Ministry of Education, Youth and Sports) funding LM 2018104, Czech Republic;
BMBF (Bundesministerium f\"{u}r Bildung und Forschung) contracts 05H12UM5, 05H15UMCNA and 05H18UMCNA, Germany;
INFN  (Istituto Nazionale di Fisica Nucleare),  Italy;
MIUR (Ministero dell'Istruzione, dell'Universit\`a e della Ricerca),  Italy;
CONACyT  (Consejo Nacional de Ciencia y Tecnolog\'{i}a),  Mexico;
IFA (Institute of Atomic Physics) Romanian 
CERN-RO No. 1/16.03.2016 
and Nucleus Programme PN 19 06 01 04,  Romania;
INR-RAS (Institute for Nuclear Research of the Russian Academy of Sciences), Moscow, Russia; 
JINR (Joint Institute for Nuclear Research), Dubna, Russia; 
NRC (National Research Center)  ``Kurchatov Institute'' and MESRF (Ministry of Education and Science of the Russian Federation), Russia; 
MESRS  (Ministry of Education, Science, Research and Sport), Slovakia; 
CERN (European Organization for Nuclear Research), Switzerland; 
STFC (Science and Technology Facilities Council), United Kingdom;
NSF (National Science Foundation) Award Numbers 1506088 and 1806430,  U.S.A.;
ERC (European Research Council)  ``UniversaLepto'' advanced grant 268062, ``KaonLepton'' starting grant 336581, Europe.

Individuals have received support from:
Charles University Research Center (UNCE/SCI/013), Czech Republic;
Ministero dell'Istruzione, dell'Universit\`a e della Ricerca (MIUR  ``Futuro in ricerca 2012''  grant RBFR12JF2Z, Project GAP), Italy;
Russian Science Foundation (RSF 19-72-10096), Russia;
the Royal Society  (grants UF100308, UF0758946), United Kingdom;
STFC (Rutherford fellowships ST/J00412X/1, ST/M005798/1), United Kingdom;
ERC (grants 268062,  336581 and  starting grant 802836 ``AxScale'');
EU Horizon 2020 (Marie Sk\l{}odowska-Curie grants 701386, 754496, 842407, 893101, 101023808).

The data used in this paper were collected before February 2022.

\bibliographystyle{na62}
\bibliography{refs}

\end{document}